\documentclass[onecolumn, amssymb, preprint, showpacs]{revtex4-1}				  % use default format for submitting

\usepackage{graphicx}			% Include figure files
\usepackage{dcolumn}			% Align table columns on decimal point
\usepackage{bm}					% bold math
\usepackage{float}			% Include float files         https://www.ctan.org/pkg/float?lang=en
\usepackage{amsmath,amsfonts}	% popular packages from the American Mathematical Society
\usepackage{url}					% https://www.ctan.org/pkg/url
\usepackage{setspace}		% Sets spacing    https://www.ctan.org/pkg/setspace

\usepackage[utf8]{inputenc}

\usepackage{newtxtext,newtxmath}

\usepackage{graphicx}                       % figures
\usepackage{geometry}                       % page layout
\usepackage{bm}            					% mathematical typesetting
\usepackage{enumitem}                       % nicer enumerated lists
\usepackage{endnotes}                       % endotes

\usepackage{subcaption}

\usepackage{psfrag}
\usepackage{units}

\usepackage[printonlyused,nolist,nohyperlinks]{acronym}
\usepackage{xspace}

\usepackage{xfrac}

\usepackage{amssymb}

\usepackage{filecontents}
\usepackage{accents}
\usepackage{esint}
\usepackage{dsfont}

\usepackage{array}
\usepackage{tabularx}
\usepackage{booktabs}
\usepackage{multirow}

\usepackage[table]{xcolor}		% must be loaded BEFORE pgfplots, as pgfplots alread loads xcolor, but without option table

\usepackage{tikz}
\usetikzlibrary{calc,trees,positioning,arrows,chains,%
	decorations,shapes,matrix,plotmarks,shadows,
	decorations.pathmorphing,decorations.pathreplacing,decorations.markings,spy}

\usepackage{pgfplots}
\usepackage{pgfplotstable}
\pgfplotsset{compat=newest, 
	colormap/bone/.style={colormap={bone}{rgb(0cm)=(0,0,0.0013021)
			rgb(1cm)=(0.0034314,0.0034314,0.0060355)
			rgb(2cm)=(0.0068627,0.0068627,0.010769)
			rgb(3cm)=(0.010294,0.010294,0.015502)
			rgb(4cm)=(0.013725,0.013725,0.020236)
			rgb(5cm)=(0.017157,0.017157,0.024969)
			rgb(6cm)=(0.020588,0.020588,0.029703)
			rgb(7cm)=(0.02402,0.02402,0.034436)
			rgb(8cm)=(0.027451,0.027451,0.03917)
			rgb(9cm)=(0.030882,0.030882,0.043903)
			rgb(10cm)=(0.034314,0.034314,0.048637)
			rgb(11cm)=(0.037745,0.037745,0.05337)
			rgb(12cm)=(0.041176,0.041176,0.058104)
			rgb(13cm)=(0.044608,0.044608,0.062837)
			rgb(14cm)=(0.048039,0.048039,0.06757)
			rgb(15cm)=(0.051471,0.051471,0.072304)
			rgb(16cm)=(0.054902,0.054902,0.077037)
			rgb(17cm)=(0.058333,0.058333,0.081771)
			rgb(18cm)=(0.061765,0.061765,0.086504)
			rgb(19cm)=(0.065196,0.065196,0.091238)
			rgb(20cm)=(0.068627,0.068627,0.095971)
			rgb(21cm)=(0.072059,0.072059,0.1007)
			rgb(22cm)=(0.07549,0.07549,0.10544)
			rgb(23cm)=(0.078922,0.078922,0.11017)
			rgb(24cm)=(0.082353,0.082353,0.11491)
			rgb(25cm)=(0.085784,0.085784,0.11964)
			rgb(26cm)=(0.089216,0.089216,0.12437)
			rgb(27cm)=(0.092647,0.092647,0.12911)
			rgb(28cm)=(0.096078,0.096078,0.13384)
			rgb(29cm)=(0.09951,0.09951,0.13857)
			rgb(30cm)=(0.10294,0.10294,0.14331)
			rgb(31cm)=(0.10637,0.10637,0.14804)
			rgb(32cm)=(0.1098,0.1098,0.15277)
			rgb(33cm)=(0.11324,0.11324,0.15751)
			rgb(34cm)=(0.11667,0.11667,0.16224)
			rgb(35cm)=(0.1201,0.1201,0.16697)
			rgb(36cm)=(0.12353,0.12353,0.17171)
			rgb(37cm)=(0.12696,0.12696,0.17644)
			rgb(38cm)=(0.13039,0.13039,0.18117)
			rgb(39cm)=(0.13382,0.13382,0.18591)
			rgb(40cm)=(0.13725,0.13725,0.19064)
			rgb(41cm)=(0.14069,0.14069,0.19537)
			rgb(42cm)=(0.14412,0.14412,0.20011)
			rgb(43cm)=(0.14755,0.14755,0.20484)
			rgb(44cm)=(0.15098,0.15098,0.20957)
			rgb(45cm)=(0.15441,0.15441,0.21431)
			rgb(46cm)=(0.15784,0.15784,0.21904)
			rgb(47cm)=(0.16127,0.16127,0.22377)
			rgb(48cm)=(0.16471,0.16471,0.22851)
			rgb(49cm)=(0.16814,0.16814,0.23324)
			rgb(50cm)=(0.17157,0.17157,0.23797)
			rgb(51cm)=(0.175,0.175,0.24271)
			rgb(52cm)=(0.17843,0.17843,0.24744)
			rgb(53cm)=(0.18186,0.18186,0.25218)
			rgb(54cm)=(0.18529,0.18529,0.25691)
			rgb(55cm)=(0.18873,0.18873,0.26164)
			rgb(56cm)=(0.19216,0.19216,0.26638)
			rgb(57cm)=(0.19559,0.19559,0.27111)
			rgb(58cm)=(0.19902,0.19902,0.27584)
			rgb(59cm)=(0.20245,0.20245,0.28058)
			rgb(60cm)=(0.20588,0.20588,0.28531)
			rgb(61cm)=(0.20931,0.20931,0.29004)
			rgb(62cm)=(0.21275,0.21275,0.29478)
			rgb(63cm)=(0.21618,0.21618,0.29951)
			rgb(64cm)=(0.21961,0.21961,0.30424)
			rgb(65cm)=(0.22304,0.22304,0.30898)
			rgb(66cm)=(0.22647,0.22647,0.31371)
			rgb(67cm)=(0.2299,0.2299,0.31844)
			rgb(68cm)=(0.23333,0.23333,0.32318)
			rgb(69cm)=(0.23676,0.23676,0.32791)
			rgb(70cm)=(0.2402,0.2402,0.33264)
			rgb(71cm)=(0.24363,0.24363,0.33738)
			rgb(72cm)=(0.24706,0.24706,0.34211)
			rgb(73cm)=(0.25049,0.25049,0.34684)
			rgb(74cm)=(0.25392,0.25392,0.35158)
			rgb(75cm)=(0.25735,0.25735,0.35631)
			rgb(76cm)=(0.26078,0.26078,0.36104)
			rgb(77cm)=(0.26422,0.26422,0.36578)
			rgb(78cm)=(0.26765,0.26765,0.37051)
			rgb(79cm)=(0.27108,0.27108,0.37525)
			rgb(80cm)=(0.27451,0.27451,0.37998)
			rgb(81cm)=(0.27794,0.27794,0.38471)
			rgb(82cm)=(0.28137,0.28137,0.38945)
			rgb(83cm)=(0.2848,0.2848,0.39418)
			rgb(84cm)=(0.28824,0.28824,0.39891)
			rgb(85cm)=(0.29167,0.29167,0.40365)
			rgb(86cm)=(0.2951,0.2951,0.40838)
			rgb(87cm)=(0.29853,0.29853,0.41311)
			rgb(88cm)=(0.30196,0.30196,0.41785)
			rgb(89cm)=(0.30539,0.30539,0.42258)
			rgb(90cm)=(0.30882,0.30882,0.42731)
			rgb(91cm)=(0.31225,0.31225,0.43205)
			rgb(92cm)=(0.31569,0.31569,0.43678)
			rgb(93cm)=(0.31912,0.31912,0.44151)
			rgb(94cm)=(0.32255,0.32255,0.44625)
			rgb(95cm)=(0.32598,0.32598,0.45098)
			rgb(96cm)=(0.32941,0.33071,0.45441)
			rgb(97cm)=(0.33284,0.33545,0.45784)
			rgb(98cm)=(0.33627,0.34018,0.46127)
			rgb(99cm)=(0.33971,0.34491,0.46471)
			rgb(100cm)=(0.34314,0.34965,0.46814)
			rgb(101cm)=(0.34657,0.35438,0.47157)
			rgb(102cm)=(0.35,0.35911,0.475)
			rgb(103cm)=(0.35343,0.36385,0.47843)
			rgb(104cm)=(0.35686,0.36858,0.48186)
			rgb(105cm)=(0.36029,0.37331,0.48529)
			rgb(106cm)=(0.36373,0.37805,0.48873)
			rgb(107cm)=(0.36716,0.38278,0.49216)
			rgb(108cm)=(0.37059,0.38752,0.49559)
			rgb(109cm)=(0.37402,0.39225,0.49902)
			rgb(110cm)=(0.37745,0.39698,0.50245)
			rgb(111cm)=(0.38088,0.40172,0.50588)
			rgb(112cm)=(0.38431,0.40645,0.50931)
			rgb(113cm)=(0.38775,0.41118,0.51275)
			rgb(114cm)=(0.39118,0.41592,0.51618)
			rgb(115cm)=(0.39461,0.42065,0.51961)
			rgb(116cm)=(0.39804,0.42538,0.52304)
			rgb(117cm)=(0.40147,0.43012,0.52647)
			rgb(118cm)=(0.4049,0.43485,0.5299)
			rgb(119cm)=(0.40833,0.43958,0.53333)
			rgb(120cm)=(0.41176,0.44432,0.53676)
			rgb(121cm)=(0.4152,0.44905,0.5402)
			rgb(122cm)=(0.41863,0.45378,0.54363)
			rgb(123cm)=(0.42206,0.45852,0.54706)
			rgb(124cm)=(0.42549,0.46325,0.55049)
			rgb(125cm)=(0.42892,0.46798,0.55392)
			rgb(126cm)=(0.43235,0.47272,0.55735)
			rgb(127cm)=(0.43578,0.47745,0.56078)
			rgb(128cm)=(0.43922,0.48218,0.56422)
			rgb(129cm)=(0.44265,0.48692,0.56765)
			rgb(130cm)=(0.44608,0.49165,0.57108)
			rgb(131cm)=(0.44951,0.49638,0.57451)
			rgb(132cm)=(0.45294,0.50112,0.57794)
			rgb(133cm)=(0.45637,0.50585,0.58137)
			rgb(134cm)=(0.4598,0.51059,0.5848)
			rgb(135cm)=(0.46324,0.51532,0.58824)
			rgb(136cm)=(0.46667,0.52005,0.59167)
			rgb(137cm)=(0.4701,0.52479,0.5951)
			rgb(138cm)=(0.47353,0.52952,0.59853)
			rgb(139cm)=(0.47696,0.53425,0.60196)
			rgb(140cm)=(0.48039,0.53899,0.60539)
			rgb(141cm)=(0.48382,0.54372,0.60882)
			rgb(142cm)=(0.48725,0.54845,0.61225)
			rgb(143cm)=(0.49069,0.55319,0.61569)
			rgb(144cm)=(0.49412,0.55792,0.61912)
			rgb(145cm)=(0.49755,0.56265,0.62255)
			rgb(146cm)=(0.50098,0.56739,0.62598)
			rgb(147cm)=(0.50441,0.57212,0.62941)
			rgb(148cm)=(0.50784,0.57685,0.63284)
			rgb(149cm)=(0.51127,0.58159,0.63627)
			rgb(150cm)=(0.51471,0.58632,0.63971)
			rgb(151cm)=(0.51814,0.59105,0.64314)
			rgb(152cm)=(0.52157,0.59579,0.64657)
			rgb(153cm)=(0.525,0.60052,0.65)
			rgb(154cm)=(0.52843,0.60525,0.65343)
			rgb(155cm)=(0.53186,0.60999,0.65686)
			rgb(156cm)=(0.53529,0.61472,0.66029)
			rgb(157cm)=(0.53873,0.61945,0.66373)
			rgb(158cm)=(0.54216,0.62419,0.66716)
			rgb(159cm)=(0.54559,0.62892,0.67059)
			rgb(160cm)=(0.54902,0.63366,0.67402)
			rgb(161cm)=(0.55245,0.63839,0.67745)
			rgb(162cm)=(0.55588,0.64312,0.68088)
			rgb(163cm)=(0.55931,0.64786,0.68431)
			rgb(164cm)=(0.56275,0.65259,0.68775)
			rgb(165cm)=(0.56618,0.65732,0.69118)
			rgb(166cm)=(0.56961,0.66206,0.69461)
			rgb(167cm)=(0.57304,0.66679,0.69804)
			rgb(168cm)=(0.57647,0.67152,0.70147)
			rgb(169cm)=(0.5799,0.67626,0.7049)
			rgb(170cm)=(0.58333,0.68099,0.70833)
			rgb(171cm)=(0.58676,0.68572,0.71176)
			rgb(172cm)=(0.5902,0.69046,0.7152)
			rgb(173cm)=(0.59363,0.69519,0.71863)
			rgb(174cm)=(0.59706,0.69992,0.72206)
			rgb(175cm)=(0.60049,0.70466,0.72549)
			rgb(176cm)=(0.60392,0.70939,0.72892)
			rgb(177cm)=(0.60735,0.71412,0.73235)
			rgb(178cm)=(0.61078,0.71886,0.73578)
			rgb(179cm)=(0.61422,0.72359,0.73922)
			rgb(180cm)=(0.61765,0.72832,0.74265)
			rgb(181cm)=(0.62108,0.73306,0.74608)
			rgb(182cm)=(0.62451,0.73779,0.74951)
			rgb(183cm)=(0.62794,0.74252,0.75294)
			rgb(184cm)=(0.63137,0.74726,0.75637)
			rgb(185cm)=(0.6348,0.75199,0.7598)
			rgb(186cm)=(0.63824,0.75672,0.76324)
			rgb(187cm)=(0.64167,0.76146,0.76667)
			rgb(188cm)=(0.6451,0.76619,0.7701)
			rgb(189cm)=(0.64853,0.77093,0.77353)
			rgb(190cm)=(0.65196,0.77566,0.77696)
			rgb(191cm)=(0.65539,0.78039,0.78039)
			rgb(192cm)=(0.66078,0.78382,0.78382)
			rgb(193cm)=(0.66616,0.78725,0.78725)
			rgb(194cm)=(0.67155,0.79069,0.79069)
			rgb(195cm)=(0.67693,0.79412,0.79412)
			rgb(196cm)=(0.68231,0.79755,0.79755)
			rgb(197cm)=(0.6877,0.80098,0.80098)
			rgb(198cm)=(0.69308,0.80441,0.80441)
			rgb(199cm)=(0.69847,0.80784,0.80784)
			rgb(200cm)=(0.70385,0.81127,0.81127)
			rgb(201cm)=(0.70924,0.81471,0.81471)
			rgb(202cm)=(0.71462,0.81814,0.81814)
			rgb(203cm)=(0.72001,0.82157,0.82157)
			rgb(204cm)=(0.72539,0.825,0.825)
			rgb(205cm)=(0.73078,0.82843,0.82843)
			rgb(206cm)=(0.73616,0.83186,0.83186)
			rgb(207cm)=(0.74154,0.83529,0.83529)
			rgb(208cm)=(0.74693,0.83873,0.83873)
			rgb(209cm)=(0.75231,0.84216,0.84216)
			rgb(210cm)=(0.7577,0.84559,0.84559)
			rgb(211cm)=(0.76308,0.84902,0.84902)
			rgb(212cm)=(0.76847,0.85245,0.85245)
			rgb(213cm)=(0.77385,0.85588,0.85588)
			rgb(214cm)=(0.77924,0.85931,0.85931)
			rgb(215cm)=(0.78462,0.86275,0.86275)
			rgb(216cm)=(0.79,0.86618,0.86618)
			rgb(217cm)=(0.79539,0.86961,0.86961)
			rgb(218cm)=(0.80077,0.87304,0.87304)
			rgb(219cm)=(0.80616,0.87647,0.87647)
			rgb(220cm)=(0.81154,0.8799,0.8799)
			rgb(221cm)=(0.81693,0.88333,0.88333)
			rgb(222cm)=(0.82231,0.88676,0.88676)
			rgb(223cm)=(0.8277,0.8902,0.8902)
			rgb(224cm)=(0.83308,0.89363,0.89363)
			rgb(225cm)=(0.83847,0.89706,0.89706)
			rgb(226cm)=(0.84385,0.90049,0.90049)
			rgb(227cm)=(0.84923,0.90392,0.90392)
			rgb(228cm)=(0.85462,0.90735,0.90735)
			rgb(229cm)=(0.86,0.91078,0.91078)
			rgb(230cm)=(0.86539,0.91422,0.91422)
			rgb(231cm)=(0.87077,0.91765,0.91765)
			rgb(232cm)=(0.87616,0.92108,0.92108)
			rgb(233cm)=(0.88154,0.92451,0.92451)
			rgb(234cm)=(0.88693,0.92794,0.92794)
			rgb(235cm)=(0.89231,0.93137,0.93137)
			rgb(236cm)=(0.89769,0.9348,0.9348)
			rgb(237cm)=(0.90308,0.93824,0.93824)
			rgb(238cm)=(0.90846,0.94167,0.94167)
			rgb(239cm)=(0.91385,0.9451,0.9451)
			rgb(240cm)=(0.91923,0.94853,0.94853)
			rgb(241cm)=(0.92462,0.95196,0.95196)
			rgb(242cm)=(0.93,0.95539,0.95539)
			rgb(243cm)=(0.93539,0.95882,0.95882)
			rgb(244cm)=(0.94077,0.96225,0.96225)
			rgb(245cm)=(0.94616,0.96569,0.96569)
			rgb(246cm)=(0.95154,0.96912,0.96912)
			rgb(247cm)=(0.95692,0.97255,0.97255)
			rgb(248cm)=(0.96231,0.97598,0.97598)
			rgb(249cm)=(0.96769,0.97941,0.97941)
			rgb(250cm)=(0.97308,0.98284,0.98284)
			rgb(251cm)=(0.97846,0.98627,0.98627)
			rgb(252cm)=(0.98385,0.98971,0.98971)
			rgb(253cm)=(0.98923,0.99314,0.99314)
			rgb(254cm)=(0.99462,0.99657,0.99657)
			rgb(255cm)=(1,1,1)}}
}

	\usepackage{cleveref}
	\creflabelformat{equation}{(#2#1#3)}
	\crefname{equation}{\hspace*{-0.5ex}}{Eqs.}
	\Crefname{equation}{Equation}{Equations}
	\crefname{chapter}{Chapter}{Chapters}
	\Crefname{chapter}{Chapter}{Chapters}
	\crefname{section}{Sec.}{Sections}
	\Crefname{section}{Section}{Sections}
	\crefname{figure}{Fig.}{Figures}
	\Crefname{figure}{Figure}{Figures}
	\crefname{appendix}{Appendix}{Appendices}
	\Crefname{appendix}{Appendix}{Appendices}
	\crefname{table}{Table}{Table}
	\Crefname{table}{Table}{Tables}
	
	\crefname{subsection}{Sec.}{Sections}
	\Crefname{subsection}{Section}{Sections}

	\makeatletter
	\renewcommand\@seccntformat[1]{\csname the#1copy\endcsname.\hskip 1em\relax}
	\makeatother
	
\begin{document}

		\renewcommand{\thefootnote}{\arabic{footnote}}
		
		\title{Broadband Multizone Sound Rendering by Jointly Optimizing the Sound Pressure and Particle Velocity}\thanks{\textit{The following article has been submitted to the Journal of the Acoustical Society of America. After it is \\[-10pt] published, it will be found at http://scitation.aip.org/JASA.}}
		
		\author{M. Buerger} \email{michael.buerger@FAU.de; Corresponding author.} 
		\author{C. Hofmann}
		\author{W. Kellermann}
		\affiliation{Chair of Multimedia Communications and Signal Processing, Friedrich-Alexander-Universität Erlangen-Nürnberg (FAU), 91058 Erlangen, Germany}
		
		% DEFINITIONS OF MICHAEL BUERGER

% GENERAL DEFINITIONS
\newcommand{\mat}[1]{\ensuremath{\mathbf{#1}}}
\newcommand{\jj}{\mathrm{i}}
\newcommand{\ie}{i.e.}
\newcommand{\eg}{e.g.}
\newcommand{\cf}{cf.~}
\newcommand{\wrt}{w.r.t.~}
\newcommand{\withoutlog}{w.\,l.\,o.\,g.\xspace}
\newcommand{\grad}{\operatorname{grad}}
\newcommand{\abs}[1]{\left\lvert#1\right\rvert}
\newcommand{\norm}[1]{\left\lVert#1\right\rVert_2}
\newcommand{\tran}{^{\operatorname{T}}}
\newcommand{\herm}{^{\operatorname{H}}}
\newcommand{\e}[1]{\operatorname{e}^{#1}}
\newcommand{\textd}{\mathrm{d}}
\newcommand{\expval}[1]{\mathcal{E}\left\{#1\right\}}

% for matlab2tikz
\newlength\figureheight
\newlength\figurewidth
\newlength\wavefieldwidth
\newcommand{\xlabelMACRO}{}
\newcommand{\xtickMACRO}{}
\newcommand{\ylabelMACRO}{}
\newcommand{\ytickMACRO}{}
\newcommand{\labelsizeMACRO}{}
\newcommand{\colorbarMACRO}{}
\newcommand{\fontsizeMACRO}{}

\newcommand{\legendone}{}
\newcommand{\legendtwo}{}
\newcommand{\legendthree}{}
\newcommand{\legendfour}{}
\newcommand{\legendfive}{}
\newcommand{\legendsix}{}

% SPECIFIC NOTATION
\newcommand{\bright}{\text{B}}
\newcommand{\dark}{\text{D}}
\newcommand{\regionofinterest}{\mathcal{R}} %^\mathrm{int}}
\newcommand{\brightzone}{\mathcal{R}^\bright}
\newcommand{\darkzone}{\mathcal{R}^\dark}
\newcommand{\volumeofinterest}{\mathcal{V}^\text{int}}
\newcommand{\Gdes}{H_\des}
\newcommand{\phisource}{\phi_\text{src}}
\newcommand{\fsample}{f_\text{s}}

\newcommand{\eei}[1]{\mat{e}_#1}
\newcommand{\des}{\text{des}}
\newcommand{\greensfunction}{G} %_\text{3D}}
\newcommand{\vecgreensfunction}{\mat{g}}
\newcommand{\vectransferfunctiondes}{\mat{h}_\text{des}^\text{p}}
\newcommand{\matgreensfunction}{\mat{G}}
\newcommand{\ggv}{\mat{g}_{\vv}}
\newcommand{\ggp}{\mat{g}_{p}}
\newcommand{\hhpressure}{\mat{g}^{\text{p}}}
\newcommand{\hhvel}{\mat{g}^{\vel}}
\newcommand{\hhveldes}{\mat{h}_\des^\vel}
\newcommand{\HHp}{\mat{H}_{pp}}
\newcommand{\kk}{\mat{k}}
\newcommand{\NC}{{N_\text{C}}}
\newcommand{\Nin}{{N}}
\newcommand{\Nout}{{M}}
\newcommand{\NL}{{N_\text{L}}}
\newcommand{\NM}{{N_\text{M}}}
\newcommand{\Ngrid}{{N_\text{grid}}}
\newcommand{\pvec}{\mat{p}}
\newcommand{\Vrad}{V_\text{rad}}
\newcommand{\Vtan}{V_\text{tan}}
\newcommand{\vel}{\text{vel}}
\newcommand{\vv}{\vec{v}}
\newcommand{\vvec}{\mat{v}}
\newcommand{\vvecrad}{\vvec_\text{rad}}
\newcommand{\vvectan}{\vvec_\text{tan}}
\newcommand{\VV}{\vec{V}}
\newcommand{\ww}{\mat{w}}
\newcommand{\Zp}{Z_\mathrm{p}}
\newcommand{\xrad}{x_\text{rad}}
\newcommand{\xtan}{x_\text{tan}}
\newcommand{\xx}{\vec{x}}
\newcommand{\xxbright}{\xx^\bright}
\newcommand{\xxdark}{\xx^\dark}
\newcommand{\xxSrc}{\vec{x}^\text{S}}
\newcommand{\xxM}{\xx^\text{M}}
\newcommand{\xxL}{\yy}
\newcommand{\xxvar}[1]{\xx_{#1}}
\newcommand{\xxrad}{\xx_\text{rad}}
\newcommand{\xxradn}{\xx_{\text{rad},n}}
\newcommand{\xxtan}{\xx_\text{tan}}
\newcommand{\xxtann}{\xx_{\text{tan},n}}
\newcommand{\xt}{\xx,t}
\newcommand{\xomega}{\xx,\omega}
\newcommand{\xin}[1]{\xx_{\tin,#1}}
\newcommand{\xout}[1]{\xx_{\tout,#1}}
\newcommand{\xoutadd}[1]{\xx_{\tout\text{\_add},#1}}
\newcommand{\xxradNULL}{\xx_{\text{rad},0}}
\newcommand{\xxtanNULL}{\xx_{\text{tan},0}}
\newcommand{\xxradONE}{\xx_{\text{rad},1}}
\newcommand{\xxtanONE}{\xx_{\text{tan},1}}
\newcommand{\xxradTWO}{\xx_{\text{rad},2}}
\newcommand{\xxtanTWO}{\xx_{\text{tan},2}}
\newcommand{\yy}{\vec{y}}
\newcommand{\Ri}{R_\text{in}}
\newcommand{\Ro}{R_\text{out}}
\newcommand{\tin}{\text{in}}
\newcommand{\tout}{\text{out}}
\newcommand{\trad}{\text{rad}}
\newcommand{\ttan}{\text{tan}}
\newcommand{\vecrenderingfilt}{\mat{w}}
\newcommand{\p}[1]{p_{#1}}
\newcommand{\pdes}[1]{p_{\text{des,}#1}}
\newcommand{\dif}[1]{\Delta p_{#1}}
\newcommand{\difdes}[1]{\Delta p_{\text{des,}#1}}

\newcommand{\circlevariable}{c}
\newcommand{\zonevariable}{z}
\newcommand{\outerpressurematrix}{\boldsymbol{\mat{E}}}
\newcommand{\differencematrix}{\mat{D}}
\newcommand{\identitymatrix}{\mat{I}}
\newcommand{\zeromatrix}{\mat{0}}

\newcommand{\calDmat}{\boldsymbol{\mathcal{D}}}

\newcommand{\VtannNC}{V_{\text{tan},\mathcal{N}}}
\newcommand{\VradnNC}{V_{\text{rad},\mathcal{N}}}

% for underbraces in matrices
\newcommand\undermat[2]{%
  \makebox[0pt][l]{$\smash{\underbrace{\phantom{%
    \begin{matrix}#2\end{matrix}}}_{#1}}$}#2}
		
% for LONG diagonal dotted lines in matrices
\newcommand{\diagdots}[3][-28]{%
  \rotatebox{#1}{\makebox[0pt]{\makebox[#2]{\xleaders\hbox{$\cdot$\hskip#3}\hfill\kern0pt}}}%
}		
		
\newcommand{\diagdotsMB}{\smash{\hspace*{10pt}\raisebox{-0.5\normalbaselineskip}{\diagdots{5em}{.25em}}}\hspace*{-10pt}}
\newcommand{\diagdotsMBshort}{\smash{\hspace*{-2pt}\raisebox{-0\normalbaselineskip}{\diagdots{3em}{.25em}}}\hspace*{-10pt}}

\definecolor{gray}{rgb}{0.5,0.5,0.5}
\newcommand{\gray}{\color{gray}}
\definecolor{darkgray}{rgb}{0.35,0.35,0.35}
\newcommand{\darkgray}{\color{darkgray}}

		% ========================= ACRONYMS
\begin{acronym}
	\acro{JPVM}{Joint Pressure and Velocity Matching}
	\acro{MSE}{Mean Squared Error}	
	\acro{PM}{Pressure Matching}	
	\acro{RIR}{Room Impulse Response}
	\acro{SNR}{Signal-to-Noise Ratio}
	\acro{WFS}{Wave Field Synthesis}
	\acro{LWE}{Loudspeaker Weight Energy}
	\acro{WNG}{White Noise Gain}
\end{acronym}

\newcommand{\JPVM}{\ac{JPVM}\xspace}
\newcommand{\JPVMmod}{JPVM+\xspace}
\newcommand{\MSE}{\ac{MSE}\xspace}
\newcommand{\PM}{\ac{PM}\xspace}
\newcommand{\RIR}{\ac{RIR}\xspace}
\newcommand{\RIRs}{\acp{RIR}\xspace}
\newcommand{\SNR}{\ac{SNR}\xspace}
\newcommand{\SNRs}{\acp{SNR}\xspace}
\newcommand{\WFS}{\ac{WFS}\xspace}
\newcommand{\LWE}{\ac{LWE}\xspace}
\newcommand{\WNG}{\ac{WNG}\xspace}

		\date{\today}

		\begin{abstract}
			In this paper, a recently proposed approach to multizone sound field synthesis, referred to as \JPVM, is investigated analytically using a spherical harmonics representation of the sound field. The approach is motivated by the Kirchhoff--Helmholtz integral equation and aims at controlling the sound field inside the local listening zones by evoking the sound pressure and particle velocity on surrounding contours. 
			Based on the findings of the modal analysis, an improved version of \JPVM is proposed which provides both better performance and lower complexity. 
			In particular, it is shown analytically that the optimization of the tangential component of the particle velocity vector, as is done in the original \JPVM approach, is very susceptible to errors and thus not pursued anymore. The analysis furthermore provides fundamental insights as to how the spherical harmonics used to describe the 3D variant sound field translate into 2D basis functions as observed on the contours surrounding the zones. 
			By means of simulations, it is verified that discarding the tangential component of the particle velocity vector ultimately leads to an improved performance. Finally, the impact of sensor noise on the reproduction performance is assessed.
		\end{abstract}

		\maketitle

% ================================================================================================

\pagebreak
% ================================================================================================

% =========================================================================
\acresetall
\section{Introduction}
% =========================================================================
Since the introduction of stereophonic sound in the 1930s, the number of loudspeakers utilized for audio reproduction has been increasing steadily. In home entertainment systems, 5.1 systems are common and have even been extended to 22.2 channels. Nowadays, modern cinemas even deploy hundreds of loudspeakers in order to create an immersive audio experience, usually via \WFS \cite{berkhout1993acoustic,spors2008theory,rabenstein2014acousticstoday}, Ambisonics \cite{gerzon1985ambisonics,ward2001reproduction,ahrens2008analytical}, or related approaches.

In recent years, much research effort has also been invested in personalized audio reproduction, \ie, synthesizing individualized acoustic scenes for multiple listeners in different areas of the reproduction room \cite{choi2002generation,poletti2008investigation,shin2010maximization,ahrens2011analytical,spors2010local,wu2011spatial,jin2013multizone,radmanesh2013generation,coleman2013optimizing,helwani2014synthesis,jin2014multizone,Cai2014sound,poletti2015approach,betlehem2015personal,buerger2015impact,buerger2015multi,poletti2016generation,winter2017time,hahn2017synthesis}. A prominent use case is given in cars, where the driver could listen to the information provided by a navigation system while all passengers can enjoy their favorite audio content. The different approaches to multizone sound reproduction vary widely: Some methods are based on \WFS and utilize analytically derived driving functions for the loudspeakers \cite{spors2010local,ahrens2011analytical,spors2011local,helwani2014synthesis}; others exploit the fact that sound fields can be described using orthogonal basis functions \cite{wu2011spatial,jin2013multizone,jin2014multizone}, such as cylindrical or spherical harmonics; there are also different multi-point approaches, where the sound pressure \cite{poletti2008investigation}, the acoustic energy/contrast between the different zones \cite{choi2002generation,shin2010maximization}, or both quantities are optimized \cite{Cai2014sound}. Multi-point techniques often use a description of the sound field in the spatial domain, but they can also be formulated in the modal domain \cite{jin2013multizone}, \ie, using basis functions.

No matter how the problem of personal audio is approached, the ultimate goal of all techniques is to evoke a desired sound pressure distribution within a certain region  of interest. All the multi-point approaches above have in common that merely the sound pressure is considered and that the control points are typically distributed in the interior of the local listening areas. For practical applications, the latter implies that real microphones (which need to be utilized in order to capture the acoustic transfer functions for all loudspeakers) obstruct the interior of the actual listening area. To mitigate this problem, the concept of \JPVM \cite{buerger2015multi} was proposed, which only requires control points on contours around the local listening areas. This can be achieved by not only optimizing the sound pressure, but also taking the particle velocity vector into account. According to the Kirchhoff--Helmholtz integral equation, it suffices to correctly evoke the sound pressure and the (radial component of) the particle velocity vector on a closed contour/surface around a source-free area/volume in order to obtain the desired sound pressure in the entire interior. 

In this contribution, \JPVM is analyzed in the modal domain, which provides insights that cannot easily be explained in the transducer domain. The findings thus obtained are exploited to enhance the performance while decreasing the computational complexity at the same time. The remainder of this document is structured as follows: First, a brief review of pressure matching \cite{poletti2008investigation} is given in \cref{sec:PM} before an improved version of \JPVM is presented in \cref{sec:JPVM_enhancement}. The analysis underlying the modifications of \JPVM is carried out in \cref{sec:analysis}, and the theoretical insights are verified by simulations in \cref{sec:experiments}, which also contains an investigation of the impact of sensor noise on the reproduction performance. Finally, the work is concluded in \cref{sec:conclusion}.

% =========================================================================
\section{Review of Pressure Matching (PM)}
\label{sec:PM}
% =========================================================================
As a starting point, we want to briefly review the concept of \PM \cite{poletti2008investigation}. For the analytic description of \PM, let $P(\xx,\omega)$ be the sound pressure at position $\xx$ in the region of interest $\regionofinterest$, and let $\omega = 2\pi f$ denote the angular frequency. The reproduced sound pressure is generated by a set of $\NL$ loudspeakers located at positions $\xxL_l$, with $l=1,\ldots,\NL$. These loudspeakers are typically modeled as infinitely long line sources (producing a height-invariant 2D sound field) or point sources (3D case). Throughout this work, we always consider point sources, the radiation characteristics of which are determined by the 3D Green's function \cite{williams1999fourier},
\begin{equation}
\greensfunction(\xx|\xxL_l,\omega) = \frac{1}{4\pi}\cdot \frac{\e{-\jj k \norm{\xxL_l - \xx}}}{\norm{\xxL_l - \xx}},
\label{eq:greens_function_3D}
\end{equation}
where $\jj$ is the imaginary unit, and the wave number $k=\omega/c$ describes the ratio between angular frequency $\omega$ and speed of sound $c$. The reproduced sound pressure at position $\xx$ can thus be expressed as
\begin{equation}
P(\xx,\omega) = \sum\limits_{l=1}^{\NL} \greensfunction(\xx|\xxL_l,\omega) \underbrace{W_l(\omega) S(\omega)}_{S_l(\omega)},
\label{eq:reproduced_pressure}
\end{equation}
with $S_l(\omega)$ being the driving signal of the $l$-th loudspeaker, which is obtained by filtering the source signal $S(\omega)$ with the corresponding loudspeaker prefilter $W_l(\omega)$. \Cref{eq:reproduced_pressure} may be written more compactly as
\begin{equation}
P(\xx,\omega) = \vecgreensfunction\tran(\xx,\omega) \vecrenderingfilt(\omega) S(\omega),
\label{eq:reproduced_pressure_vector_notation}	
\end{equation}
where the superscript $(\cdot)\tran$ indicates vector/matrix transposition, the column vector $\vecrenderingfilt(\omega) = [W_1(\omega),\ldots,W_\NL(\omega)]\tran$ contains the frequency responses of the loudspeaker prefilters, and the 3D Green's functions for all loudspeakers are captured in $\vecgreensfunction(\xx,\omega) = [\greensfunction(\xx|\xxL_1,\omega),\ldots,\greensfunction(\xx|\xxL_\NL,\omega)]\tran$.

In a multizone scenario, the region of interest $\regionofinterest$ accommodates several local listening areas: a `bright zone' $\brightzone$, in which one or more virtual sound sources shall be synthesized, and possibly multiple `dark zones' $\darkzone$ where the acoustic scene synthesized in the bright zone shall not be perceivable. For the sake of simplicity and without loss of generality, we restrict ourselves to a single virtual source to be synthesized in the bright zone $\brightzone$ and a single dark zone $\darkzone$ throughout this document. Let us describe the desired sound pressure in $\regionofinterest$ according to $P_\des(\xx,\omega)=\Gdes(\xx,\omega) S(\omega)$, where $\Gdes(\xx,\omega)$ denotes the target transfer function at position $\xx \in \regionofinterest $. In the bright zone $\brightzone$, the desired transfer function $\Gdes$ preferably represents an elementary solution of the acoustic wave equation, such as a cylindrical wave, a spherical wave, a plane wave, or a superposition thereof. For the dark zone, the ultimate goal is $\Gdes(\xx,\omega)=0 \ \forall \xx\in\darkzone$ such that no acoustic energy is radiated into it. Following the approach by Poletti \cite{poletti2008investigation}, the reproduced pressure field is matched with the desired one at a set of $\NM$ discrete positions, referred to as matching points or control points. The corresponding problem formulation for pressure matching is given by
\begin{equation}
\min\limits_{\mat{w(\omega)}}  \norm{\matgreensfunction(\omega) \mat{w(\omega)} - \vectransferfunctiondes(\omega)}^2,
\label{eq:pressure_matching_aim}
\end{equation}
where $\vectransferfunctiondes(\omega) = [\Gdes(\xx_1,\omega),\ldots,\Gdes(\xx_\NM,\omega)]\tran$ is a column-vector capturing the desired transfer functions from the virtual target source to the control points in both the bright and the dark zone, and the transfer functions from the loudspeakers to the control points are accommodated in matrix  $\matgreensfunction(\omega) = [\vecgreensfunction(\xx_1,\omega),\ldots,\vecgreensfunction(\xx_\NM,\omega)]\tran$.

Given that there are no loudspeakers or physical barriers between the different local listening areas, it is physically not feasible to have perfect silence in one area and a desired pressure distribution in another area. This implies that the target sound field can only be correctly reproduced at a limited number of positions, but not in a spatial continuum. If the number of loudspeakers is smaller than the number of control points, an approximate solution of \cref{eq:pressure_matching_aim} is required, which can be obtained, for example, using Singular Value Decomposition (SVD) or the Moore--Penrose pseudoinverse with Tikhonov regularization \cite{golub96matrix}. Additional constraints can also be incorporated, \eg, in order to limit the array effort or \LWE \cite{betlehem2012sound}. It is also possible to impose a constraint on the acoustic contrast between the control points in the individual zones \cite{Cai2014sound}. Alternative approaches, such as LASSO \cite{lilis2010sound,radmanesh2013generation}, additionally minimize the $L_1$-norm of $\mat{w}$ in order to control the number of active loudspeakers and, thus, obtain a sparse solution.

% =========================================================================
\section{Enhancement of Joint Pressure and Velocity Matching (JPVM)}
\label{sec:JPVM_enhancement}
% =========================================================================
In this section, we present an improved version of the original \JPVM approach \cite{buerger2015multi}, which will be referred to as \JPVMmod. Before the optimization problem for \JPVMmod is formulated in \cref{sec:JPVMmod_cost_function}, the underlying representation of the particle velocity vector is discussed in \cref{sec:particle_velocity_vector}. For simplicity, we assume that both the loudspeakers and the local listening areas are located in the $x$-$y$-plane, \ie, the target sound field is defined for a 2D plane, while the loudspeakers are still modeled as point sources. The extension to three dimensions is not addressed here for the sake of brevity. However, the necessary incorporation of the $z$-component of the particle velocity vector, which is required to optimize the sound field on a sphere rather than on a circular contour, can be done analogously to the $x$- and $y$-components. 

% -------------------------------------------------------------------------
\subsection{The Particle Velocity Vector}
\label{sec:particle_velocity_vector}
\begin{figure}
	\includegraphics[width=8cm]{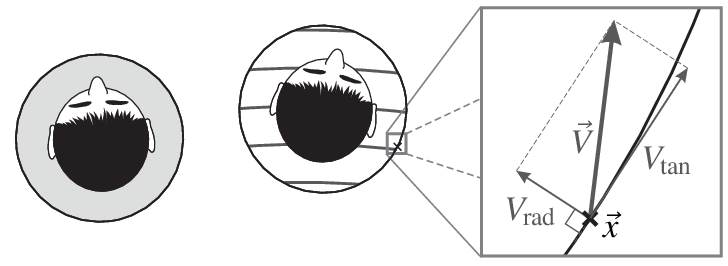}
	\caption{Radial and tangential component of the particle velocity vector defined on the circular contour around a local listening area.}
	\label{fig:def_particle_velocity}
\end{figure}

% -------------------------------------------------------------------------
Below, we revisit how the particle velocity vector is approximated and incorporated into the optimization problem for the original \JPVM approach \cite{buerger2015multi}. The particle velocity vector $\VV(\xx, \omega)$ is, apart from the scalar sound pressure $P(\xx, \omega)$, the second important quantity to characterize sound waves. With time-dependency $\e{\jj\omega t}$, a relation between the two quantities can be established by Euler's equation \cite{beranek1954acoustics},
\begin{equation}
-\grad  P(\xx,\omega) =  \jj \omega \rho \VV(\xx, \omega),
\label{eq:euler_frequ_domain}
\end{equation}
where $\rho$ is the density of the propagation medium. The components of the particle velocity vector can be defined for arbitrary coordinate systems, \eg, along the $x$- and $y$-axis. In the original \JPVM approach \cite{buerger2015multi}, the vector is composed of the radial and tangential component on a circular contour around each local listening area, \ie, around the bright zone $\brightzone$ and dark zone $\darkzone$, as illustrated in \cref{fig:def_particle_velocity}. Here, we are only interested in the radial component $\Vrad(\xx, \omega)$, as the tangential component can be discarded (see \cref{subsec:modal_representation_tangential_component,subsec:controllability_P_V} for the theoretical motivation and \cref{subsec:exp_impact_of_tan_component} for simulation results). 

\begin{figure}
	\includegraphics{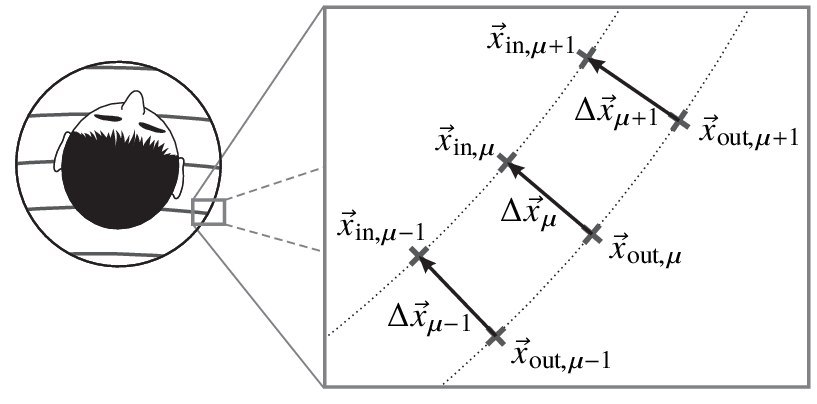}
	\caption{Pairs of control points located on the contour around a local listening area for \JPVMmod such that the radial component of the particle velocity vector can be approximated.} 
	\label{fig:control_points_definition}% 
\end{figure}

Due to the complexity of real sound fields, \cref{eq:euler_frequ_domain} can generally not be evaluated analytically. In practice, the particle velocity can be obtained using a B-Format microphone \cite{craven1977coincident}, which provides the sound pressure at its center as well as the three components of the particle velocity vector. Alternatively, these signals can be obtained using an omnidirectional microphone as well as three figure-of-eight microphones arranged orthogonally to each other, which is referred to as native B-Format \cite{benjamin2005native}. If not all components of the particle velocity vector are required, fewer figure-of-eight microphones are obviously sufficient. Nevertheless, B-format and directional microphones are usually much more expensive and bulky than omnidirectional microphones. Therefore, we approximate the particle velocity vector in this work by spatially sampling the sound field at the control points with omnidirectional microphones and by subsequently computing the spatial difference quotient. For this purpose, the control points are arranged pairwise on two concentric circles of slightly different radii, as illustrated in \cref{fig:control_points_definition}. The inner and outer circle accommodate $M$ control points each, which are located at positions $\xin{\mu}$ and $\xout{\mu}$, respectively, with $\mu=1,\ldots,M$. The radial component of the particle velocity vector may then be approximated according to 
\begin{equation}
\Vrad(\xout{\mu},\omega) \approx -\frac{1}{\jj \omega \rho} \frac{ P(\xin{\mu},\omega) - P(\xout{\mu},\omega) }{ \norm{\Delta\xx_\mu}},
\label{eq:approx_rad_component}
\end{equation}
where $\Delta\xx_\mu =  \xin{\mu} - \xout{\mu}$. For a sound reproduction system, the evoked radial component of the particle velocity can be obtained by inserting \cref{eq:reproduced_pressure_vector_notation} into \cref{eq:approx_rad_component}, which yields 
\begin{equation}
\Vrad(\xout{\mu},\omega) \approx \frac{ \vecrenderingfilt\tran(\omega) \left(\vecgreensfunction(\xin{\mu},\omega) - \vecgreensfunction(\xout{\mu},\omega) \right) S(\omega) }{ - \jj \omega \rho \norm{\Delta\xx_\mu}}.
\label{eq:approx_rad_component_reproduced}
\end{equation}
That is, the components of the approximated particle velocity vector can be expressed in terms of loudspeaker prefilters $\vecrenderingfilt$ and acoustic transfer functions $\vecgreensfunction$. 

\begin{figure}
	\includegraphics{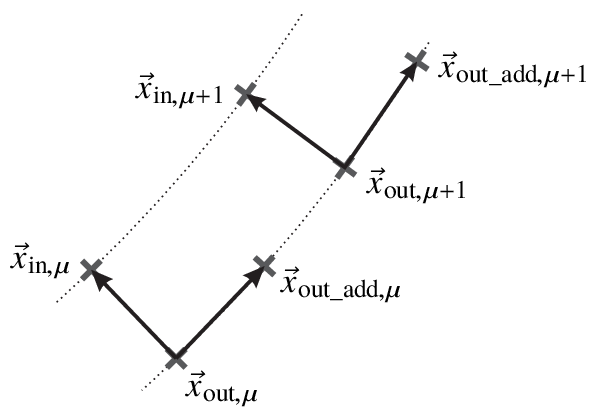}
	\caption{L-shaped groups of control points located on the contour around a local listening as required for the original version of JPVM \cite{buerger2015multi}.}
	\label{fig:control_points_L_shape}% 
\end{figure}

The tangential component of the particle velocity vector could be described analogously by computing the pressure difference along the angular direction, as is done in the original contribution \cite{buerger2015multi}. For this purpose, additional microphones need to be placed at positions $\xoutadd{\mu}$ on the outer circle, as illustrated in \cref{fig:control_points_L_shape}, where groups of three microphones are arranged in an L-shape so as to approximate both components of the particle velocity vector. The tangential component could then be approximated according to 
\begin{equation}
\Vtan(\xout{\mu},\omega) \approx -\frac{1}{\jj \omega \rho} \frac{ P(\xoutadd{\mu},\omega) - P(\xout{\mu},\omega) }{ \norm{ \xoutadd{\mu+1} - \xoutadd{\mu} }},
\label{eq:approx_tan_component}
\end{equation}
which can again be expressed in terms of $\vecrenderingfilt$ and $\vecgreensfunction$. However, as already mentioned above, only the radial component is of interest here, and \cref{eq:approx_tan_component} is included for completeness. 

From \cref{eq:approx_rad_component,eq:approx_rad_component_reproduced,eq:approx_tan_component} it is obvious that the spacing between the control points plays a crucial role, as is always the case when utilizing differential microphone arrays \cite{teutsch2001first}: Large spacings result in a low spatial aliasing frequency, whereas systems with a small spacing are more prone to sensor noise, which must be taken into account for practical realizations. Therefore, the noise susceptibility of \JPVMmod is analyzed in \cref{sec:sensor_noise}.

% -------------------------------------------------------------------------
\subsection{Joint Optimization of Pressure and Particle Velocity}
\label{sec:JPVMmod_cost_function}
% -------------------------------------------------------------------------

We now derive the optimization problem for the improved version of \JPVM, referred to as \JPVMmod, where the individual steps of the derivation follow the same line as the ones presented in the original paper \cite{buerger2015multi}. To formally describe the cost function for \JPVMmod, we first arrange the acoustic transfer functions from the $l$-th loudspeaker to all $M$ control points on each circle in column vectors $\vecgreensfunction_{\circlevariable,l}^\zonevariable(\omega)=\left[\greensfunction(\xx_{\circlevariable,1}^\zonevariable|\xxL_l,\omega), \ldots, \greensfunction(\xx_{\circlevariable,M}^\zonevariable|\xxL_l,\omega) \right]\tran$, where the superscript $\zonevariable\in\{\bright,\dark\}$ indicates the bright or dark zone, and the subscript $\circlevariable\in\{\tin,\tout\}$ refers to the inner or outer circle of control points. Using this notation, the transfer functions from the entire set of loudspeakers to all control points can be conveniently captured in
\begin{equation}
\matgreensfunction(\omega) = 
\begin{bmatrix} 
\vecgreensfunction_{\tout,1}^\bright(\omega), \ldots, \vecgreensfunction_{\tout,\NL}^\bright(\omega) \\[3pt]
\vecgreensfunction_{\tin,1}^\bright(\omega), \ldots, \vecgreensfunction_{\tin,\NL}^\bright(\omega) \\[3pt]
\vecgreensfunction_{\tout,1}^\dark(\omega), \ldots, \vecgreensfunction_{\tout,\NL}^\dark(\omega) \\[3pt]
\vecgreensfunction_{\tin,1}^\dark(\omega), \ldots, \vecgreensfunction_{\tin,\NL}^\dark(\omega) \\[3pt]
\end{bmatrix},	
\end{equation}
such that the sound pressure at the control points can be compactly written as
\begin{equation}
\pvec(\omega) = \underbrace{ \matgreensfunction(\omega) \vecrenderingfilt(\omega)}_{\hhpressure(\omega)} S(\omega),
\label{eq:true_outer_pressure}
\end{equation}
with $\hhpressure$ representing the transfer function of the overall system \wrt the sound pressure.

\newcommand{\underbracedmatrixleft}[2]{%
	\left[\;
	\smash[b]{\underbrace{
			\begin{matrix}#1\end{matrix}
		}_{#2}}
	\right.
	\vphantom{\underbrace{\begin{matrix}#1\end{matrix}}_{#2}}
}
\newcommand{\underbracedmatrixright}[2]{%
	\left.
	\smash[b]{\underbrace{
			\begin{matrix}#1\end{matrix}
		}_{#2}}
	\;\right]
	\vphantom{\underbrace{\begin{matrix}#1\end{matrix}}_{#2}}
}

As only the radial component of the particle velocity vector is considered here, which is motivated in \cref{sec:analysis}, the original difference matrix \cite{buerger2015multi} used for computing the corresponding pressure differences simplifies to 
\begin{equation}
\differencematrix(\omega) = -\frac{1}{\jj \omega \rho \Delta R} 
\underbracedmatrixleft{
	-\identitymatrix 	& \identitymatrix  \\ 
	\zeromatrix 	& \zeromatrix    \\
}{\parbox{.8cm}{\scriptsize\centering bright zone}}
\underbracedmatrixright{
	\zeromatrix  		& \zeromatrix \\ 
	-\identitymatrix & \identitymatrix  \\
}{\parbox{.8cm}{\scriptsize\centering dark zone}}
\label{eq:difference matrix}
\end{equation}
where $\Delta R = \norm{\Delta\xx_\mu}\ \forall \mu$ represents the absolute difference between radii $R_\tin$ and $R_\tout$  of the inner and the outer circle of control points, respectively. Furthermore, $\identitymatrix$ and $\zeromatrix$ denote an identity and zero matrix, respectively, of dimensions $M \times M$. The resulting radial components of the particle velocity vector along the contours of the bright and dark zone are then given by
\begin{equation}
\vvec_\trad(\omega) = \underbrace{\differencematrix(\omega) \matgreensfunction(\omega) \vecrenderingfilt(\omega)}_{\hhvel(\omega)} S(\omega),
\label{eq:true_rad_vel}
\end{equation}
where $\hhvel$ represents the transfer function of the overall system \wrt the radial component of the particle velocity vector.

Using \cref{eq:true_rad_vel} and \cref{eq:true_outer_pressure}, a weighted least squares optimization criterion for both quantities can be formulated, 
\begin{equation}
\min\limits_{\mat{w}(\omega)} \left\{ \kappa(\omega) \norm{ \matgreensfunction(\omega)\mat{w}(\omega) - \vectransferfunctiondes(\omega) }^2 	\hspace{5pt} + (1-\kappa(\omega)) \norm{ \differencematrix(\omega)\matgreensfunction(\omega)\mat{w}(\omega) - \hhveldes(\omega) }^2
\right\},\hspace{-5pt}
\label{eq:pressure_and_velocity_minimization_problem}
\end{equation}
where $\vectransferfunctiondes$ and $\hhveldes$ represent the desired pressure and velocity transfer functions, respectively, and $\kappa(\omega) \in [0,1]$ is used to adjust the relative weight of each quantity. Similar to the original \JPVM approach \cite{buerger2015multi}, the optimization problem  \cref{eq:pressure_and_velocity_minimization_problem} can be reformulated using stacked matrices such that a common least squares problem is obtained, which can then be solved in a closed form, \eg, using a Moore--Penrose pseudoinverse with Tikhonov regularization \cite{golub96matrix}. For the sake of brevity, these steps are omitted here. Note that the cost function for \JPVMmod does not contain the tangential component of the particle velocity vector implying that the optimization problem has a lower complexity compared to \JPVM.

%\acresetall
% =========================================================================
\section{Analysis of the Impact of the Particle Velocity Vector Components on JPVM}
\label{sec:analysis}
% =========================================================================
In this section, we want to analyze how the individual components of the particle velocity vector affect the behavior and performance of \JPVM. For this purpose, the sound pressure is formulated in the modal domain, where important fundamental relations are discussed in \cref{subsec:modal_representation_sound_field}. Similar to the sound pressure, a modal representation of the radial and tangential component of the particle velocity vector is introduced in \cref{subsec:modal_representation_radial_component} and \cref{subsec:modal_representation_tangential_component}, respectively. Finally, a representative example is given in \cref{subsec:controllability_P_V} which illustrates how well the different quantities on the contour are suited to control the sound field in the interior.

% -------------------------------------------------------------------------
\subsection{Modal Representation of Sound Fields Evoked by Point Sources in a 2D Plane}
\label{subsec:modal_representation_sound_field}
% -------------------------------------------------------------------------

The loudspeakers in this work are modeled as point sources, which is a simple yet reasonable approximation of real loudspeakers. This implies that, even though the previous considerations are limited to a 2D plane, the sound fields are in fact three-dimensional, where any point $\xx$ in the 3D space can be addressed by its azimuth angle $\phi$, colatitude angle $\theta$, and radius $r$. It is well known that arbitrary 3D sound fields within a source-free region can be described as \cite{williams1999fourier}
\begin{equation}
P(r,\theta,\phi,\omega) = \sum\limits_{n=0}^\infty \sum\limits_{m=-n}^n \alpha_{mn}(\omega) j_n(kr) Y_n^m(\theta,\phi),
\label{eq:pressure_spherical_harmonics}
\end{equation}
where 
\begin{equation}
Y_n^m(\theta,\phi) = \underbrace{\sqrt{ \frac{(2n+1)}{4\pi}  \frac{(n-\abs{m})!}{(n+\abs{m})!}   }}_{b_{mn}} P_n^{\abs{m}}(\cos \theta) \e{+\jj m \phi}
\label{eq:spherical_harmonics}
\end{equation}
are the spherical harmonics of order $n$ and degree $m$, with $P_n^m$ being the associated Legendre functions, $j_n$ is the $n$-th order spherical Bessel function, and $\alpha_{mn}$ are the modal weights.

\begin{figure}
	\centering
	\includegraphics[width=7cm]{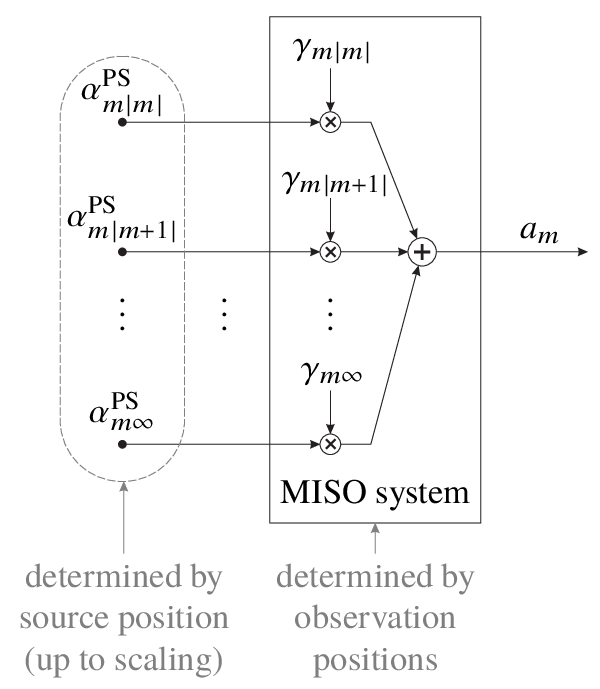}
	\caption{Schematic illustration of the mapping of the modal weights $\alpha_{mn}^\text{PS}$, with $n=\abs{m},\ldots,\infty$, onto the Fourier coefficients $a_m$ representing the sound pressure on the contour around a local listening area. For brevity, the frequency-dependency is omitted here. }
	\label{fig:mapping_alpha_a}
\end{figure}

For a point source with frequency response $A(\omega)$ located at radius $r_0$ and direction ($\theta_0, \phi_0$), the modal weights are given by \cite{colton2012inverse}
\begin{equation}
\alpha_{mn}^\text{PS}(\omega) = - \jj A(\omega) k h_n^{(2)}(kr_0) b_{mn} P_n^{\abs{m}}(\cos \theta_0) \e{-\jj m\phi_0},
\label{eq:modal_weights_point_source}
\end{equation}
where $h_n^{(2)}$ denotes the $n$-th order spherical Hankel function of the second kind. Note that \cref{eq:pressure_spherical_harmonics} with modal weights according to \cref{eq:modal_weights_point_source} and $A(\omega)=1 \ \forall \omega$ is equivalent to the 3D Green's function \cref{eq:greens_function_3D}.  As we assume in this work that the loudspeakers and observation points lie in the $x$-$y$ plane, \ie, $\theta_0 = \pi / 2$ and $\theta = \pi / 2$, we omit the colatitude angles $\theta$ and $\theta_0$ in the notation for brevity. Using the above modal representation, the sound pressure evoked by a single loudspeaker, which is modeled as an ideal point source here, can be expressed as 
\begin{equation}
P(r,\phi,\omega) = \sum\limits_{n=0}^\infty \sum\limits_{m=-n}^n \alpha_{mn}^\text{PS}(\omega) j_n(kr) b_{mn} P_n^{\abs{m}}(1) \e{+\jj m \phi}.
\label{eq:sound_pressure_point_source_plane}
\end{equation}
On the other hand, the sound pressure at the circular contour around a local listening area can be represented by a Fourier series. More precisely, we express the sound pressure at the outer circle of control points with radius $R_\tout$ as
\begin{equation}
P(R_\tout,\phi,\omega)	= \sum\limits_{\mu=-\infty}^{\infty} a_\mu(R_\tout,\omega)  \e{\jj \mu \phi},
\label{eq:fourier_representation_sound_pressure_3D}
\end{equation}
where $a_\mu$ are the Fourier coefficients. The coefficients $a_\mu$ obtained for a single point source can be computed by equating \cref{eq:fourier_representation_sound_pressure_3D} with \cref{eq:sound_pressure_point_source_plane} and evaluating the inner product with the complex harmonics, \ie, 
\begin{equation}
a_\mu(R_\tout,\omega) = \frac{1}{2 \pi} \int\limits_{0}^{2\pi} \sum\limits_{n=0}^\infty \sum\limits_{m=-n}^n \alpha_{mn}^\text{PS}(\omega) j_n(k R_\tout) b_{mn} P_n^{\abs{m}}(1) \e{+\jj m \phi} \e{-\jj \mu \phi} \text{d} \phi.
\label{eq:determination_fourier_coefficient_sound_pressure_3D}
\end{equation}
Due to the orthogonality of complex harmonics, the integral may only be non-zero for $\mu = m$ such that the Fourier coefficients are given by
\begin{equation}
a_m(R_\tout,\omega) = \sum\limits_{n=\abs{m}}^\infty \alpha_{mn}^\text{PS}(\omega) \underbrace{j_n(k R_\tout) b_{mn} P_n^{\abs{m}}(1)}_{\gamma_{mn}(k, R_\tout)}.
\label{eq:fourier_coefficient_sound_pressure_3D}
\end{equation}
Note that the summation over $n$ in \cref{eq:fourier_coefficient_sound_pressure_3D} must be limited to $n\geq \abs{m}$, since $\abs{m}$ of an associated Legendre function cannot be larger $n$. This is equivalent to the limitation of the sum over $m$ from $-n,\ldots,n$ in \cref{eq:pressure_spherical_harmonics}. From \cref{eq:fourier_coefficient_sound_pressure_3D}, it can be seen that a particular Fourier coefficient $a_m$ corresponds to a weighted sum of the modal weights $\alpha_{m n}^\text{PS}$ for a given degree $m$ and all orders $n$, where the individual weights are given by a set of scalar factors $\gamma_{mn}$. This set of frequency-dependent factors $ \gamma_{m n} $ can be interpreted as a Multiple-Input Single-Output (MISO) system, which maps all modal weights for a particular index $m$ onto a Fourier coefficient representing a corresponding quantity on the contour (here: sound pressure), as illustrated in \cref{fig:mapping_alpha_a}. Thus, a particular Fourier coefficient $a_m$ obtained at the output of such a MISO system describes how well the weighted sum of all modes of degree $m$ can be observed when evaluating the sound pressure on the contour. In case the mapping due to $\gamma_{m n}$ results in a very small absolute value $ a_m$, a given set of modal weights $\alpha_{mn}^\text{PS}$ thus appears on the contour as strongly attenuated.

\begin{figure}%
	\centering
	\includegraphics[width=13cm]{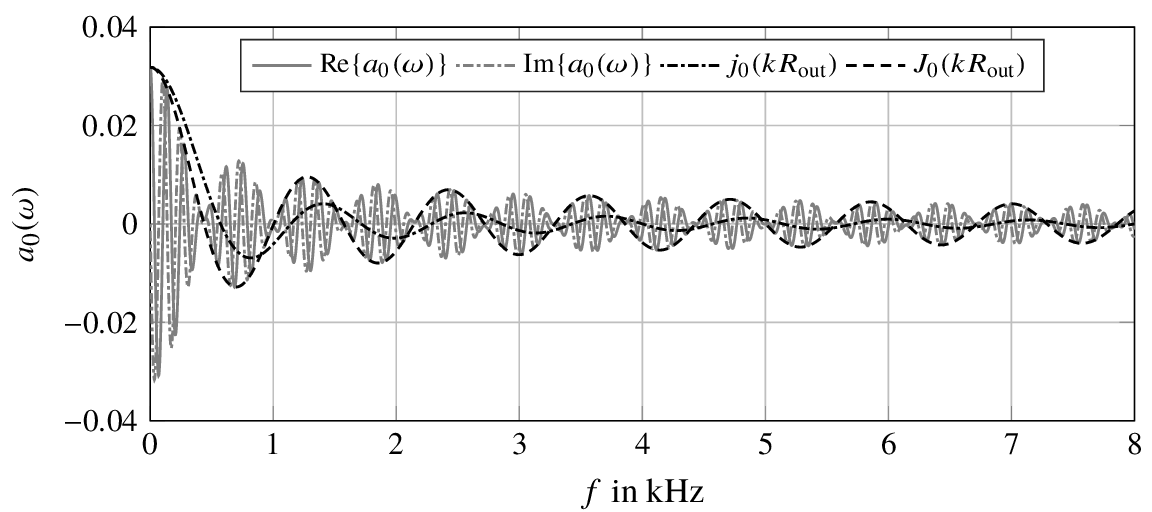}
	\caption{Fourier coefficient $a_0$ of the sound pressure $P$ at $R_\tout=\unit[0.3]{m}$ as resulting from a point source located at $r_0 = \unit[2.5]{m}$ in the $x$-$y$ plane.}%
	\label{fig:pressure_point_source_frequency}%
\end{figure}

To illustrate the mapping performed by \cref{eq:fourier_coefficient_sound_pressure_3D}, the Fourier coefficient $a_0$ is plotted in \cref{fig:pressure_point_source_frequency} for $R_\tout=\unit[0.3]{m}$ and a point source located at $r_0 = \unit[2.5]{m}$, where the source direction $\phi_0$ is irrelevant here due to $m=0$. As expected, the magnitude of $a_m$ is strongly frequency-dependent, and the weighted sum of all modes for a given $m$ cannot be observed well at certain frequencies. Interestingly, the envelope/magnitude of $\alpha_0$ very closely follows the zeroth-order \textit{cylindrical} Bessel function $J_0$ rather than the  \textit{spherical} Bessel function $j_0$, even though the sound field is resulting from a point source. This behavior cannot only be observed for the coefficient $a_0$, but for $a_m$ in general. It should be noted, however, that the cylindrical Bessel functions exhibit zero transitions, whereas the magnitude of $a_m$ does not take on the value zero for the corresponding arguments.

The previous considerations describe the mapping from given modal weights $\alpha_{mn}^\text{PS}$ onto the Fourier coefficients $a_m$ representing the sound pressure on the contour around a local listening area. If, on the other hand, the sound field shall be controlled by evoking the sound pressure on the contour, the inverse mapping comes into play. From a system-theoretical point of view, this may seem problematic, since it suggests the inversion of a MISO system. However, the factors $\gamma_{m n}$ of this MISO system only depend on the positions of the control points and are thus known. Furthermore, the modal coefficients $\alpha_{mn}^\text{PS}$ are dictated by the physics of a point source and its position. This implies that, for the given setup with a point source located in the $x$-$y$ plane, there is only a single possible set of modal coefficients (up to a scaling due to $A(\omega)$) for a given sound pressure on the contour. Only in case $a_m$ was equal to zero, infinitely many sets of modal coefficients would be possible, which is referred to as the nonuniqueness problem \cite{fazi2012nonuniqueness}. Nevertheless, the inverse mapping may be ill-conditioned \cite{fazi2007ill} and involve a strong amplification, such that low absolute values of the sound pressure on the contour may correspond to large values of the sound pressure in the interior of the local listening area. Accordingly, the sound pressure on the contour must then be evoked very precisely in order to reproduce the desired sound field properly, and even small errors on the contour may lead to a substantial deterioration in the interior. The observability of the modes on the contour can therefore be regarded as a robustness indicator: The larger the values of $a_m$ are, the less the sound field in the interior of the zone is affected by reproduction errors on the contour. 

Similarly to the sound pressure itself, the radial and tangential pressure differences on the contours around the local listening areas may also be represented in terms of Fourier series, as shown below. The expressions thus obtained provide insights into the behavior of \JPVM and serve as a basis for the modifications underlying the improved version \JPVMmod.

% -------------------------------------------------------------------------
\subsection{The Radial Component of the Particle Velocity Vector}
\label{subsec:modal_representation_radial_component}
% -------------------------------------------------------------------------

Analogously to the above, let us again consider a single loudspeaker located in the $x$-$y$ plane. Using the spherical harmonics representation of \cref{eq:pressure_spherical_harmonics,eq:spherical_harmonics,eq:modal_weights_point_source}, the radial pressure difference in the numerator of \cref{eq:approx_rad_component} for a certain angle $\phi$ on the contour can be expressed as
\begin{equation} 
\begin{split}
\Delta P_\trad(R_\tin,R_\tout,\phi,\omega) 	=& P(R_\tin,\phi,\omega) - P(R_\tout, \phi,\omega) \\
=& \sum\limits_{n=0}^\infty \sum\limits_{m=-n}^n
\alpha_{mn}^\text{PS}(\omega)  \underbrace{\left( j_n(kR_\tin) - j_n(kR_\tout) \right) b_{m n} P_n^{\abs{m}}(1)}_{\gamma_{\trad,mn}(R_\tin, R_\tout, \omega)} 
\e{\jj m \phi} \\
=& \sum\limits_{m=-\infty}^\infty \underbrace{ \sum\limits_{n=\abs{m}}^\infty 
	\alpha_{mn}^\text{PS}(\omega)  \gamma_{\trad,mn}(R_\tin, R_\tout, \omega) }_{\Delta a_{\trad,m}(R_\tin, R_\tout, \omega)}
\e{\jj m \phi},
\label{eq:rad_pressure_diff}
\end{split}
\end{equation}
where the order of the summations over $m$ and $n$ is exchanged in the last step and the limits are adapted accordingly. Similarly to $\gamma_{m n}$ in \cref{eq:fourier_coefficient_sound_pressure_3D}, the set of frequency-dependent factors $ \gamma_{\trad,m n} $ in \cref{eq:rad_pressure_diff} also describe a MISO system. The difference to \cref{eq:fourier_coefficient_sound_pressure_3D} is that the modal weights $\alpha_{m n}^\text{PS}$ are now mapped onto Fourier coefficients $\Delta a_{\trad,m}$ which represent the radial pressure difference on the contour rather than the sound pressure itself. Accordingly, the output $\Delta a_{\trad,m}$ of the system describes how well the weighted sum of all modes of degree $m$ can be observed on the contour when evaluating the radial pressure difference. In turn, it also tells us how robustly a desired sound field can be reproduced by evoking the radial pressure difference on the contour. Before discussing the behavior of the radial pressure difference, we first want to also express the tangential pressure difference, which is also optimized in the original version of \JPVM \cite{buerger2015multi}, in the same way.

% -------------------------------------------------------------------------
\subsection{The Tangential Component of the Particle Velocity Vector}
\label{subsec:modal_representation_tangential_component}
% -------------------------------------------------------------------------

Expressing the tangential pressure difference in the numerator of \cref{eq:approx_tan_component} by means of the spherical harmonics representation yields 
\begin{equation}
\begin{split}
\Delta P_\ttan(R_\tout, \phi,\omega) =& P(R_\tout,\phi+\Delta\phi,\omega) - P(R_\tout,\phi,\omega) \\
=& \sum\limits_{n=0}^\infty \sum\limits_{m=-n}^n
\alpha_{mn}^\text{PS}(\omega) \underbrace{ j_n(kR_\tout) b_{m n} P_n^{\abs{m}}(1)  \left( \e{\jj m \Delta\phi }  - 1 \right)  }_{\gamma_{\ttan,mn}(R_\tout, \Delta\phi, \omega)} 
\e{\jj m \phi} \\
=& \sum\limits_{m=-\infty}^\infty \underbrace{ \sum\limits_{n=\abs{m}}^\infty 
	\alpha_{mn}^\text{PS}(\omega)  \gamma_{\ttan,mn}(R_\tout, \Delta\phi, \omega) }_{\Delta a_{\ttan,m}(R_\tout, \Delta\phi, \omega)}
\e{\jj m \phi},
\label{eq:tan_difference}
\end{split}
\end{equation}
where $\Delta\phi$ denotes the angle between two neighboring control points on the circle with radius $R_\tout$. The frequency-dependent gain factors $\gamma_{\ttan,mn}$ represent another MISO system, which maps the modal coefficients $\alpha_{mn}^\text{PS}$ onto Fourier coefficients $ \Delta a_{\ttan,m} $ representing the tangential pressure difference along the contour. Again, these coefficients indicate how well a sound field can be controlled by evoking the tangential pressure difference on the contour. It can be seen directly from \cref{eq:tan_difference} that the tangential pressure difference component $\Delta a_{\ttan,m}$ is a scaled version of the sound pressure component $a_m$, where the scaling factor $\text{exp}\left(\jj m \Delta\phi\right)-1$ is absolutely much smaller than one for small $\Delta\phi$ and low degrees $m$. This implies that the sound pressure itself can typically be better observed than the tangential pressure difference. 

A more detailed discussion and comparison of the different mappings of modal coefficients due to $\gamma_{mn}$, $\gamma_{\trad, mn}$, and $\gamma_{\ttan,mn}$ onto the corresponding Fourier coefficients representing the sound pressure on the contour, the radial pressure difference, and the tangential pressure difference, respectively, is provided below.

% -------------------------------------------------------------------------
\subsection{Controllability of the Sound Pressure and its Radial and Tangential Difference}
\label{subsec:controllability_P_V} 
% -------------------------------------------------------------------------

As an example, let us consider a point source located at $r_0 = \unit[2.5]{m}$ and $\phi_0 = \pi$, where the circles accommodating the control points are of radii $R_\tin = \unit[0.275]{m}$ and $R_\tout = \unit[0.3]{m}$, and the observation angle is $\phi=0^\circ$. The angular difference $\Delta\phi$ is chosen such that the absolute angular distance between two control points on the outer circle is identical to the absolute radial distance $\Delta R = \unit[2.5]{cm}$. 

\begin{figure}%
	\centering
	\includegraphics[width=12.3cm]{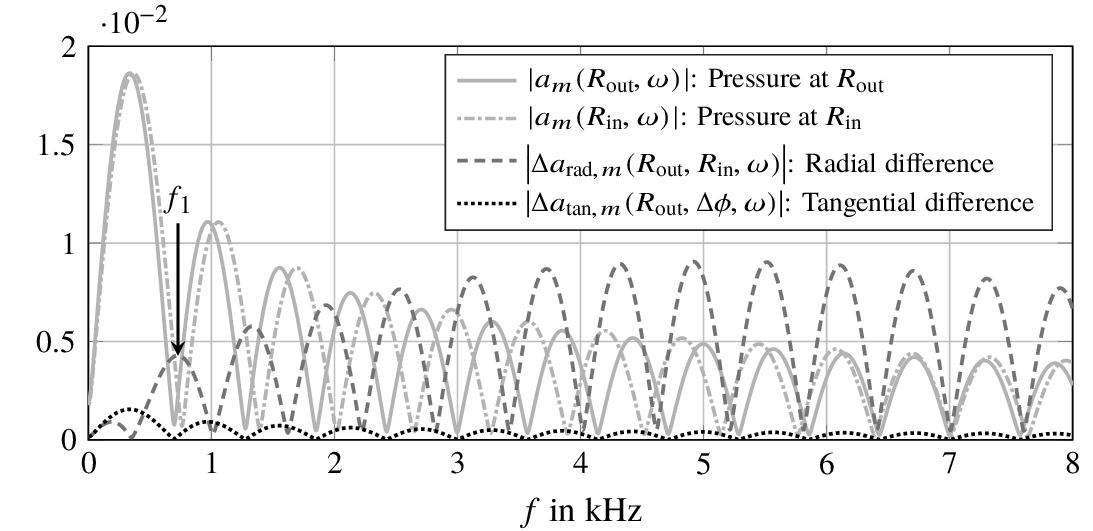}
	\caption{Magnitudes of the observed Fourier coefficients representing the sound pressure components for $m=1$ on the contour as well as the radial and tangential pressure differences.}%
	\label{fig:observed_magnitudes}%
\end{figure}

\Cref{fig:observed_magnitudes} shows the resulting Fourier coefficients $a_m$, $\Delta a_{\trad,m}$, and $\Delta a_{\ttan,m}$ for $m=1$, which represent the respective component of the sound pressure, the radial pressure difference, and the tangential pressure difference on the contour. It can be seen that sound pressure component $a_1$, whose magnitude behaves similarly as the magnitude of the Bessel function $J_1$, cannot be observed well on the contour for certain frequencies. Conversely, reproducing a sound field by controlling the sound pressure on the contour is prone to reproduction errors for these frequencies. Especially in a multizone scenario, reproduction errors are inevitably occurring as the optimization problem is typically ill-conditioned, such that a regularization is required in order to limit the gains of the loudspeaker prefilters. Therefore, a low performance must be expected for these frequencies.

A similar reasoning holds for the radial pressure difference $a_{\trad,m}$, which too exhibits low values at some frequencies, as can also be seen in \cref{fig:observed_magnitudes} for $m=1$. However, the frequencies at which the local minima occur are shifted relative to the ones of the sound pressure itself. As a consequence, the radial pressure difference has generally large values if the pressure itself is low, and vice versa, such that at least one quantity can be used to robustly excite the sound field. Only for few frequencies beyond $\unit[5]{kHz}$, the local minima of the sound pressure component and the radial pressure difference lie closely together. It is worth noting here that the location of the local minima of the radial pressure difference can be controlled by the distance $\Delta R$ between the two circles of control points. If $\Delta R \rightarrow 0$, as an extreme case, the difference becomes proportional to the spatial derivative with its extremal values being at those frequencies where the magnitude of the sound pressure component has its steepest slope, \ie, in the vicinity of the local minima. Even though this may seem to be the best choice, the absolute values of the radial pressure difference $a_{\trad,m}$ would then be very small for all frequencies which again implies a low robustness against reproduction errors. Therefore, a compromise needs to be found between the absolute values of $a_{\trad,m}$ and the locations of the local minima by choosing proper values for $\Delta R$.

Finally, the magnitude of the Fourier coefficient $\Delta a_{\ttan,m}$ describing the tangential pressure difference on the contour is also plotted in \cref{fig:observed_magnitudes} for $m=1$. In contrast to the radial pressure difference, the tangential pressure difference defined in \cref{eq:tan_difference} typically is very small. The main reason for it is that the scaling factor $\text{exp}(\jj m \Delta\phi)-1$ generally takes on very low absolute values, unless very high degrees $m$ or a large spacing $\Delta\phi$ are considered. Due to the fact that $\Delta\phi$ must be chosen sufficiently small in order to approximate the spatial derivative well, the scaling factor will be generally much smaller than one. The components $\Delta a_{\ttan,m}$ representing the tangential pressure difference are therefore also much smaller than the components $a_m$ for the pressure itself. Moreover, the local minima of $\Delta a_{\ttan,m}$ coincide with the local minima of $a_m$, such that an optimization of the tangential pressure difference does not provide any benefit over the sound pressure itself. In fact, evoking the tangential pressure difference may even degrade the reproduction performance, as shown in \cref{subsec:exp_impact_of_tan_component}.

The analysis of the above example shows that it is beneficial to simultaneously control the sound pressure and the radial pressure difference. However, one may ask whether it is also sufficient to optimize only the sound pressure itself on two different radii, which should also (implicitly) control the radial pressure difference. The answer to that question can also be found in \cref{fig:observed_magnitudes}: For a frequency of $f_1=\unit[728]{Hz}$, as an example, the magnitude of $a_1$ representing sound pressure components for $m=1$ on the contour is equally small for both radii $R_\tin$ and $R_\tout$, \ie, a local minimum of $\max \left\{ \abs{a_1(R_\tin)}, \abs{a_1(R_\tout)} \right\}$ is obtained. In fact, the \PM approach exhibits a local performance minimum for frequencies around $f_1$, as will be shown by simulation results in \cref{subsec:performance_comparision}. An explicit optimization of the radial pressure difference, on the other hand, does not suffer from this kind of performance drop, because the magnitude of the radial pressure difference component $\Delta a_{\trad,1}$ is significantly larger than the corresponding magnitudes of the individual sound pressure components for the considered frequency. In other words, the performance of the \PM approach is dictated by two pressure components of the same magnitude, whereas the \JPVM/\JPVMmod approach can also exploit and optimize a quantity whose absolute value is significantly larger due to the phase differences between the individual pressure components. Therefore, an explicit optimization of the radial pressure difference is more robust to errors than an implicit optimization. Nevertheless, the optimization of the sound pressure $a_m$ on two circles with different radii rather than a single circle is still beneficial for frequencies where the absolute value of the radial pressure difference $a_{\trad,m}$ is smaller than the absolute values of both individual sound pressures themselves.

% =========================================================================
\section{Simulations}
\label{sec:experiments}
% =========================================================================
The reproduction performance of \PM, \JPVM, and \JPVMmod is now evaluated for a simulated free-field environment, where the simulation setup and the utilized performance measures are explained in \cref{subsec:exp_setup}. Afterwards, a comparison of the three approaches is conducted in \cref{subsec:performance_comparision}, before evaluation results are shown in \cref{subsec:exp_impact_of_tan_component} which illustrate how the optimization of tangential pressure difference affects the reproduction performance. The impact of an explicit optimization of the radial component of the pressure difference is shown in \cref{subsec:exp_impact_of_rad_component} by means of synthesized sound fields, which also demonstrate the broadband behavior of \JPVMmod. Finally, it is investigated in \cref{sec:sensor_noise} how inherent noise of the microphones, which are required to capture the \RIRs in practice, affects the reproduction performance.

% -------------------------------------------------------------------------
\subsection{Simulation Setup}
\label{subsec:exp_setup}
% -------------------------------------------------------------------------

\begin{figure}%
	\centering
	\includegraphics{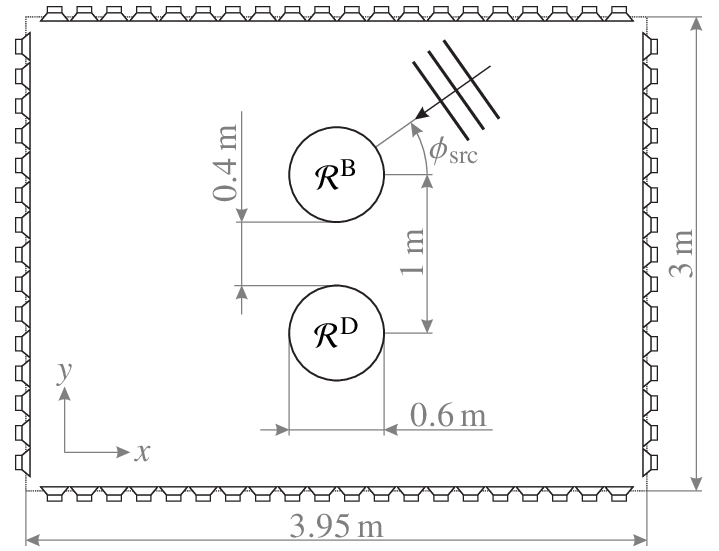}
	\caption{Schematic illustration of the reproduction setup with a plane-wave front originating from direction $\phisource$ relative to the center of the bright zone $\brightzone$.}%
	\label{fig:setup}%
\end{figure}

For a validation of the analytical findings presented above, we simulate a 70-element rectangular loudspeaker array of dimensions $\unit[3.95]{m} \times \unit[3]{m}$ with an inter-element spacing of $\unit[9.6]{cm}$, where the loudspeakers are modeled as ideal point sources. This setup allows for a direct comparison with the results of the original \JPVM paper\cite{buerger2015multi}. Two local listening areas are considered, whose centers are separated by a distance of $\unit[1]{m}$ along the $y$-axis and placed symmetrically \wrt the array center, as illustrated in \cref{fig:setup}. Each zone has a diameter of $\unit[0.6]{m}$ (implying that $R_\tout = \unit[0.3]{m}$), \ie, the edges of the bright zone $\brightzone$ and the dark zone $\darkzone$ are separated by $\unit[0.4]{m}$. The radius of the inner circle of control points is chosen as $R_\tin = \unit[0.275]{m}$, which amounts to a radial spacing of $ \Delta R= \unit[2.5]{cm}$. To allow for a fair performance comparison, the number of utilized control points is identical for all approaches and amounts to 48 per zone. That is, 24 pairs of control points are uniformly distributed around each local listening area when applying the improved version \JPVMmod and \PM. In case of the original version of \JPVM, which requires more control points on the outer circle, 16 L-shaped groups consisting of three control points each (\cf \cref{fig:control_points_L_shape}) are utilized for each zone, where both the radial and tangential spacing is $\unit[2.5]{cm}$. The relative weight $\kappa$ for the sound pressure and particle velocity in \cref{eq:pressure_and_velocity_minimization_problem} is set to 0.04, and the minimization problem is solved using the Moore--Penrose pseudioinverse, where the regularization parameter is iteratively increased until the Loudspeaker Weight Energy (LWE) \cite{betlehem2012sound} falls below the specified upper limit of $10/\NL$. As shown in the appendix, the chosen constraint on the \LWE is comparable to a \WNG of $10\log_{10}(0.1\NL)$. For comparison, a WNG of $10\log_{10}(\NL)$ corresponds to a delay-and-sum beamformer \cite{vantrees2004detection}, which is the most robust beamformer and $\unit[10]{dB}$ larger than the one obtained here. The target sound field in the upper, bright zone is given by a plane wave originating from direction $\phisource=-50^\circ$. This constitutes an especially difficult scenario as the effective distance between the zones as `observed' from the source direction $\phisource$ is close to zero. In other words, the acoustic energy needs to be focused very sharply and its level needs to rapidly drop in the lateral direction in order not to travel through the dark zone. Note that the plane wave front defining the target sound field in the bright zone does not have a magnitude of one, 
but it is scaled such that it corresponds to the magnitude which would be obtained with a point source located at a distance of $\bar{r}\approx \unit[2]{m}$ from the center of $\brightzone$, where $\bar{r}$ is the average distance from all loudspeakers to the zone center. If this scaling was not applied, the prefilters would need to compensate for the distance-dependent attenuation of the spherical waves emitted by the loudspeakers and, thus, have an inappropriately high energy. Furthermore, the above choice allows for directly relating the \LWE to the \WNG \cite{cox1986practical} (see Appendix). All simulations are conducted using a sampling frequency of $\fsample = \unit[8]{kHz}$, and a filter length $L=256$ is chosen for the loudspeaker prefilters.

As a first objective performance measure for the different reproduction algorithms, the \MSE of the overall system transfer function \wrt the sound pressure is evaluated at $\Ngrid=441$ grid points in each zone, 
\begin{equation}
\text{MSE}^{\left\{\bright, \dark \right\}}(\omega) = \frac{1}{\Ngrid} \sum\limits_{q=1}^{\Ngrid} \abs{\Gdes\left(\xx^{\left\{\bright, \dark \right\}}_q,\omega\right) - \vecgreensfunction\tran\left(\xx^{\left\{\bright, \dark \right\}}_q,\omega\right) \vecrenderingfilt(\omega) }^2,
\label{eq:MSE}
\end{equation}
where $\xxbright_q \in \brightzone$, $\xxdark_q \in \darkzone$, and the $x$- and $y$-spacing between the evaluation grid positions is $\unit[2]{cm}$. 
Furthermore, we evaluate the average level difference $\Delta L$ between the bright zone and the dark zone,
\begin{equation}
\Delta L(\omega) = \unit[10 \log_{10} \left( \frac{ E^\bright(\omega) }{ E^\dark(\omega) } \right)]{dB},
\label{eq:level_difference}
\end{equation}
where the average energies for $\brightzone$ and $\darkzone$ are computed as
\begin{equation}
E^{\left\{\bright,\dark\right\}}(\omega) = \frac{1}{\Ngrid} \sum\limits_{q=1}^{\Ngrid} \abs{\vecgreensfunction\tran\left(\xx^{\left\{\bright,\dark\right\}}_q,\omega\right) \vecrenderingfilt(\omega) }^2
\end{equation}
at the same grid positions.

% -------------------------------------------------------------------------
\subsection{Performance Comparison for \PM, \JPVM, and \JPVMmod}
\label{subsec:performance_comparision}
% -------------------------------------------------------------------------

As a first investigation, we want to compare the performance obtained with \PM \cite{poletti2008investigation}, the original \JPVM approach \cite{buerger2015multi}, and the modified version \JPVMmod presented here. As mentioned above, L-shaped groups of control points are used for \JPVM, whereas the control points are arranged in pairs in case of \PM and \JPVMmod. Even though this implies that the setup is not identical for all approaches, the number of utilized control points is the same such that the comparison is still fair. 

\begin{figure}%
	\centering	
	\includegraphics[width=15cm]{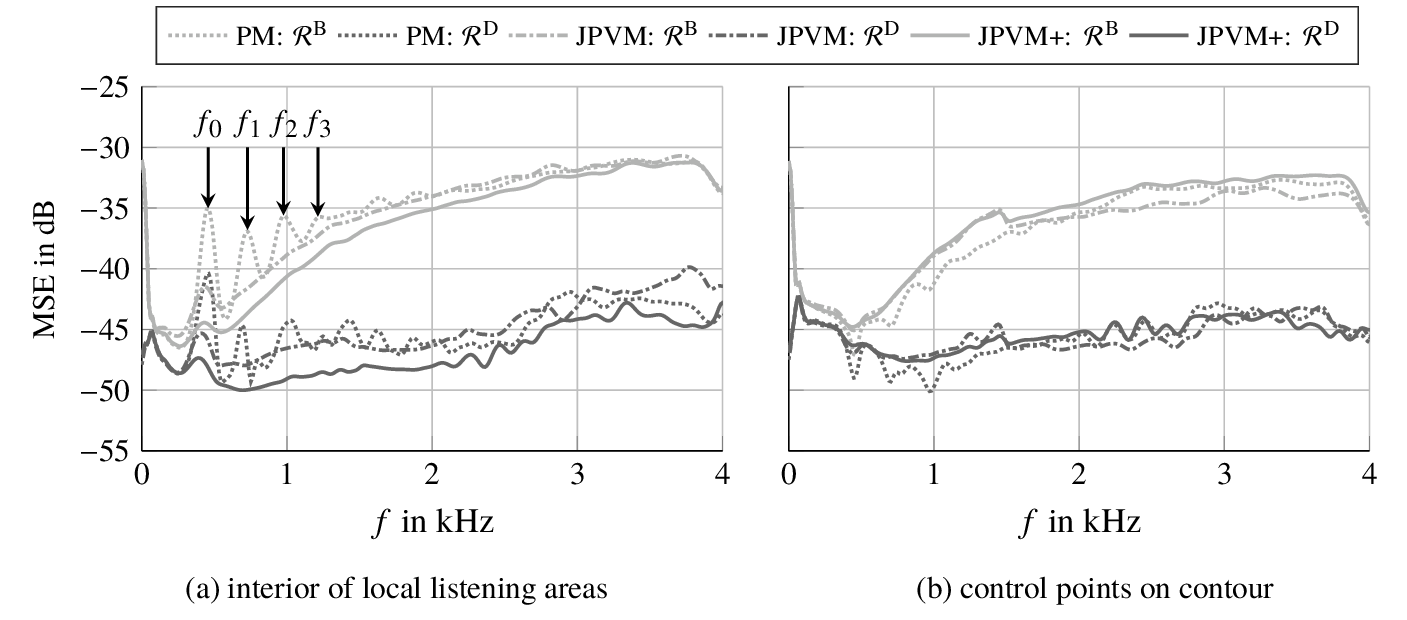}
	\caption{MSE inside (a) and at the control points around (b) the bright zone $\brightzone$ and dark zone $\darkzone$ for PM (dotted line), JPVM (dash-dotted line), and \JPVMmod (solid line).}%
	\label{fig:MSE_inside_and_contour}%
\end{figure}

The \MSE inside the local listening areas obtained for the three approaches is shown as a function of frequency in \cref{fig:MSE_inside_and_contour}(a). As can be seen, the performance of \PM oscillates very strongly in the frequency range between about $\unit[250]{Hz}$ and $\unit[1250]{Hz}$, which is attributed to the ill-conditioning stemming from the low observability of certain modes at particular frequencies (\cf \cref{sec:analysis}). For comparison purposes, the frequencies $f_m$ corresponding to the respective first local minimum of the magnitude of Fourier coefficient $a_m$ are indicated with arrows for $m=0,\ldots,3$. As can be seen, these frequencies match the local \MSE maxima very well. When optimizing the entire particle velocity vector in addition to the sound pressure, as is the case in \JPVM, the oscillation of the MSE curve can be significantly reduced, which implies a much more uniform performance and suggests less coloration of the perceived signal. However, the \MSE can be reduced even further for almost all frequencies when utilizing \JPVMmod. This is due to the fact that the optimization of the tangential pressure difference in \JPVM is very prone to (even tiny) errors, which cannot be avoided in practice. Especially in the bright zone, the performance towards higher frequencies clearly decreases for all approaches. Additional investigations showed that this can be mitigated by reducing the spacing between the loudspeakers, which leads to a lower spatial aliasing frequency, or by relaxing the constraint on the \LWE. However, higher values of the \LWE imply a lower robustness against errors typically occurring in practice, such as transducer noise, positioning errors, or variations of the properties of real loudspeakers.

\setlength{\figurewidth}{.43\textwidth}
\setlength{\figureheight}{.25\textwidth}

For completeness, the \MSE for the control points themselves (\ie, the contour) is also shown in \cref{fig:MSE_inside_and_contour}(b). Note that the \MSE values are again computed according to \cref{eq:MSE}. That is, the \MSE only captures the sound pressure and not the particle velocity vector. As expected, the \MSE is lowest in case of \PM, where all degrees of freedom are used to optimize the sound pressure at the control points. Interestingly, the \MSE at the control points has local minima at frequencies where the \MSE inside the local listening area exhibits local maxima. This again indicates the problem that a low observability of the sound pressures on the contours, which form the basis for the optimization problem of \PM, may lead to large reproduction errors in the interior of the zone. Evoking the sound pressure on the contour is thus not sufficient to infer the sound field inside the contour. In case of \JPVM and \JPVMmod, the error of the sound pressure at the control points for frequencies below about $\unit[1.5]{kHz}$ is larger compared to \PM, as some degrees of freedom are used to optimize the particle velocity (vector). Nevertheless, they result in a better performance inside the listening areas, as shown in \cref{fig:MSE_inside_and_contour}(a).

\begin{figure}%
	\centering	
	\includegraphics[width=8cm]{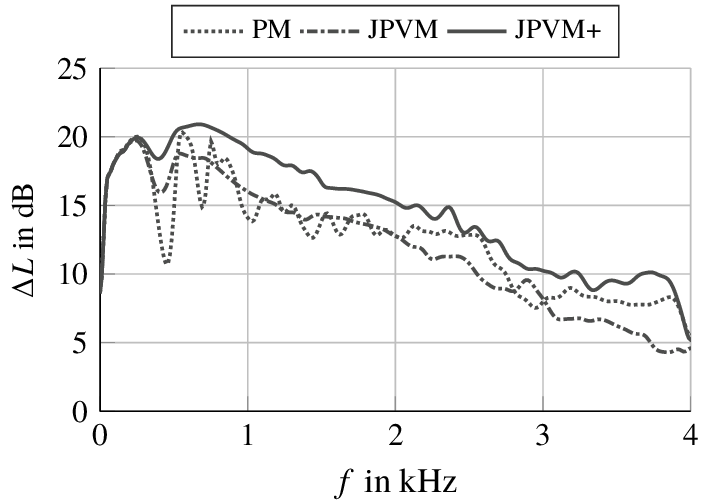}
	\caption{Level difference $\Delta L$ between the bright zone $\brightzone$ and the dark zone $\darkzone$ for PM (dotted line), JPVM (dash-dotted line), and \JPVMmod (solid line).}%
	\label{fig:level_differences}%
\end{figure}

\Cref{fig:MSE_inside_and_contour}(a) may suggest that the benefit of \JPVMmod over \PM is rather low as the absolute \MSE values in both zones are very low for all approaches. 
However, it is not sufficient to only assess the \MSE, but it is also necessary to evaluate the achieved level difference $\Delta L$ between the two local listening areas, which is illustrated in \cref{fig:level_differences}. Again, the \PM approach results in strong oscillations of the performance measure, where the level difference drops from $\unit[20]{dB}$ to almost $\unit[10]{dB}$ when increasing the frequency from $\unit[250]{Hz}$ to $\unit[450]{Hz}$. This implies that the acoustic scene to be synthesized in the bright zone does not leak into the dark zone equally strong for all frequencies, but certain frequencies produce more leakage than others. The level difference achieved with \JPVMmod, in contrast, is much more uniform and also clearly larger in the entire frequency range. This indicates that not only the overall energy in the dark zone is lower in case of \JPVMmod, but the arriving signals also have less coloration. The original \JPVM approach ranges mostly between \PM and \JPVMmod, and even though the curve is comparably smooth, the absolute level differences in certain frequency ranges are even lower than those of \PM.

\setlength{\figurewidth}{.43\textwidth}
\setlength{\figureheight}{.25\textwidth}

% -------------------------------------------------------------------------
\subsection{Impact of the Tangential Component of the Particle Velocity Vector}
\label{subsec:exp_impact_of_tan_component} 
% -------------------------------------------------------------------------

\begin{figure}%
	\centering	
	\includegraphics[width=12cm]{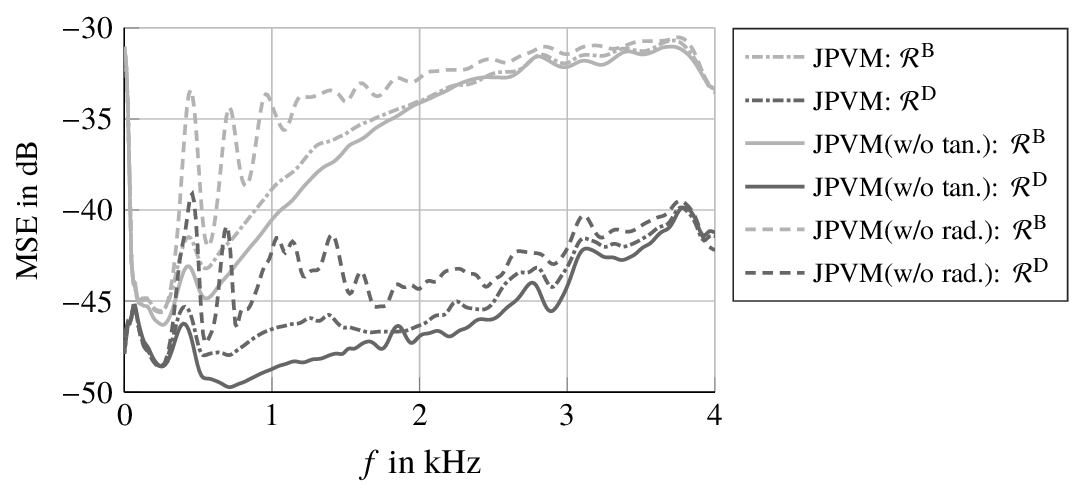}
	\caption{MSE inside the bright zone $\brightzone$ and dark zone $\darkzone$ for the original JPVM approach (dash-dotted line), the JPVM approach without the optimization of the tangential component of the particle velocity vector (solid line), and the JPVM approach without the optimization of the radial component of the particle velocity vector (dashed line).}%
	\label{fig:MSE_inside_JPVM_vs_JPVM_rad_only}%
\end{figure}

As the control points for \JPVM and \JPVMmod need to be arranged differently, we now want to evaluate a single setup and investigate in an isolated manner how the optimization of the tangential particle velocity vector component impairs the reproduction performance. For this purpose, we again consider the already evaluated setup for \JPVM with 16 L-shaped groups of control points being distributed on the contour around each zone. The resulting frequency-dependent \MSE obtained with \JPVM is again shown in \cref{fig:MSE_inside_JPVM_vs_JPVM_rad_only}, which also contains the \MSE obtained when excluding the tangential component of the particle velocity vector, \ie, when optimizing the sound pressure at all available control points and the radial component only. Note that the latter case does not correspond to \JPVMmod, as the control points are still arranged in L-shaped groups, and the number of control points on the outer circle is twice as large as for the inner circle. As can be seen in \cref{fig:MSE_inside_JPVM_vs_JPVM_rad_only}, the \MSE is almost consistently lower if the tangential component is not optimized. Only for very few frequencies, a negligible performance degradation can be observed. These results confirm the insights obtained by the theoretical analysis presented in  \cref{subsec:modal_representation_tangential_component,subsec:controllability_P_V}, and they show that the tangential component of the particle velocity vector should not be optimized, as doing so may generally reduce the reproduction performance. 
For completeness, the \MSE obtained when excluding the radial component, \ie, when only optimizing the sound pressure at all available control points and the tangential component only, is also plotted in \cref{fig:MSE_inside_JPVM_vs_JPVM_rad_only}. This is to show that the \MSE reduction obtained when neglecting the tangential component does not stem from the reduced complexity of the resulting optimization problem, and it furthermore confirms the efficacy of optimizing the radial component.

\setlength{\figurewidth}{.43\textwidth}
\setlength{\figureheight}{.25\textwidth}

% -------------------------------------------------------------------------
\subsection{Impact of the Radial Component of the Particle Velocity Vector}
\label{subsec:exp_impact_of_rad_component} 
% -------------------------------------------------------------------------

In addition to the objective performance measure evaluated above, we now want to illustrate the benefit of explicitly optimizing the radial component of the particle velocity vector by means of visualizations of the sound fields synthesized by \PM and \JPVMmod. Note that the illustrated sound fields are normalized in order to allow for a better assessment. 

\begin{figure}
	\includegraphics[width=15cm]{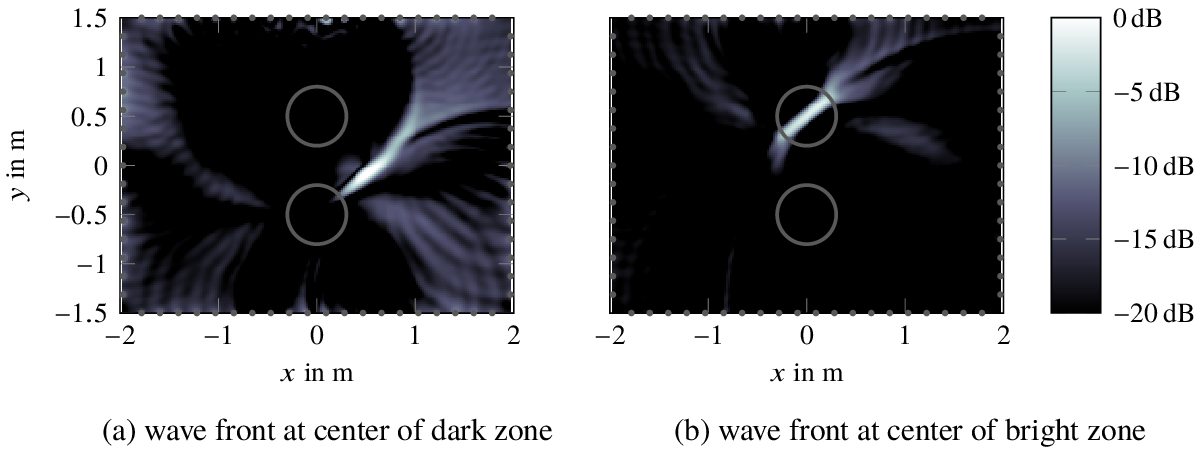}
	\caption{Reproduction of a plane-wave front for two different time instants using \JPVMmod.}
	\label{fig:wave_field_JPVM}
\end{figure}

\begin{figure}
	\centering
	\includegraphics[width=15cm]{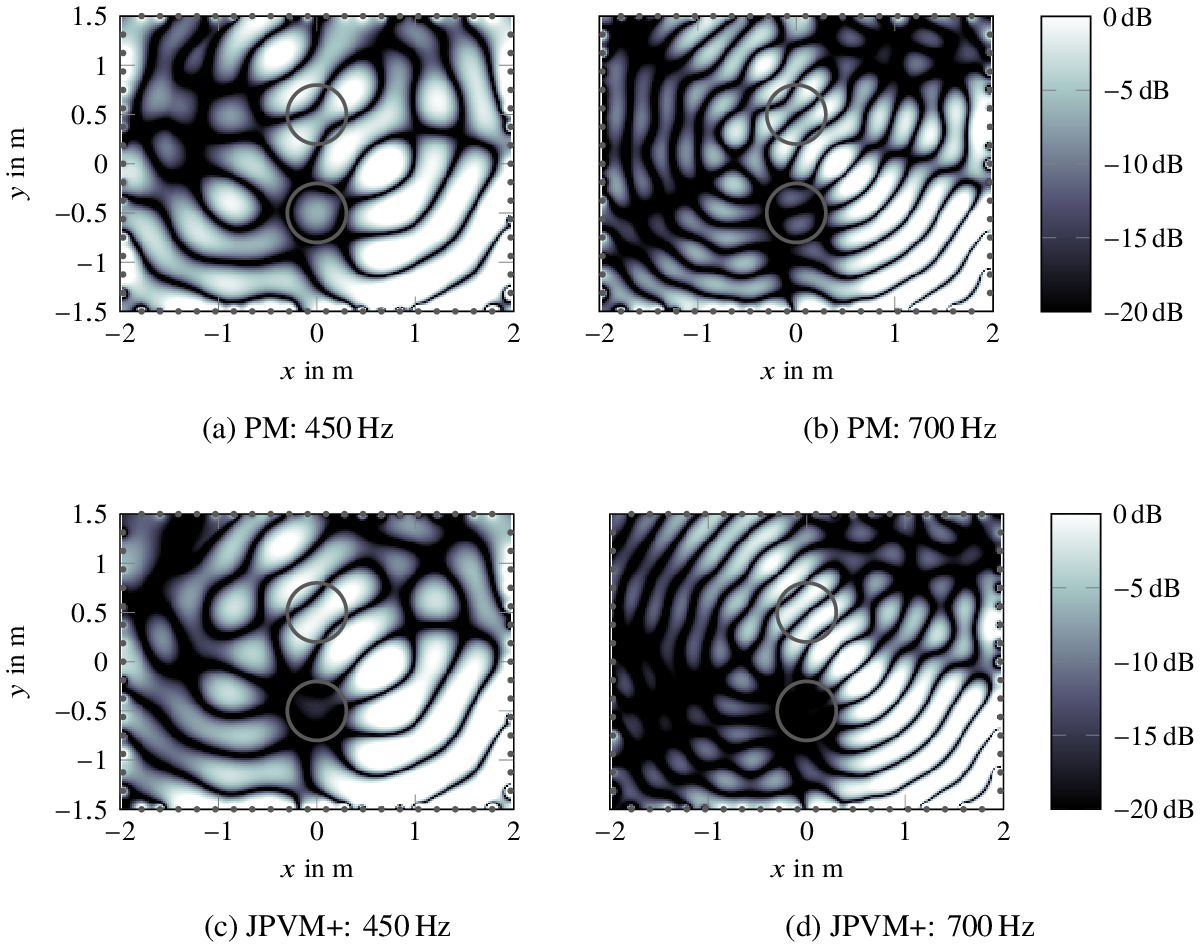}
	\caption{Reproduction of a plane-wave front of frequency $\unit[450]{Hz}$ and $\unit[700]{Hz}$ using PM and \JPVMmod.}
	\label{fig:wave_field_PM_vs_JPVM}
\end{figure}

To demonstrate the broadband behavior of \JPVMmod, a \textit{von Hann} impulse synthesized as a plane-wave front is shown in \cref{fig:wave_field_JPVM} for two time instants. The time instants are chosen such that a virtual plane perpendicular to the propagation direction of the wave front is at the center of the respective local listening area. It can be seen that the acoustic wave travels nicely around the lower, dark zone, where only the edge of the wave front leaks slightly into the interior of the zone (a). At the center of the upper, bright zone (b), a plane-wave front of desired orientation is synthesized. We do not show the corresponding plots of \PM here, as the differences for these two time instants are rather hard to see when a short impulse is reproduced. Instead, the sound fields are illustrated for a frequency of $\unit[450]{Hz}$ and $\unit[700]{Hz}$ in \cref{fig:wave_field_PM_vs_JPVM}. These frequencies are close to $f_0$ and $f_1$, which correspond to local minima of the magnitudes of the Fourier coefficients $a_0$ and $a_1$, respectively, representing the respective component of the sound pressure on the contour (see \cref{eq:fourier_coefficient_sound_pressure_3D}). Especially in the dark zone, it can be seen that the modes for $m=0$ (a) and $m=1$ (b) are erroneously excited in the interior of the zones when using \PM. That is, a large sound energy is present in the dark zone at these frequencies, whereas the desired wave front in the bright zone is clearly distorted. The \JPVMmod approach, on the other hand, only slightly excites the modes for $m=0$ in the dark zone (see \cref{fig:wave_field_PM_vs_JPVM}(c)), and the modes for $m=1$ cannot be observed at all (see \cref{fig:wave_field_PM_vs_JPVM}(d)). Similarly, the wave front in the bright zone is much less distorted for both frequencies.

% -------------------------------------------------------------------------
\subsection{Sensor Noise}
\label{sec:sensor_noise}
% -------------------------------------------------------------------------

All previous simulations are based on the free-field assumption and it is assumed that the acoustic transfer functions are perfectly known. In practice, however, the reproduction system will typically be installed in a closed room and the sound waves emitted by the loudspeakers undergo reflections. In order to compensate for these undesired reflections, the acoustic properties of the room need to be captured, possibly in an adaptive manner. The topic of listening room equalization \cite{goetze2008multi,schneider2012adaptive,schneider2012iterative,talagala2014efficient,hofmann2016higher} is not in the scope of this work and we would like to refer to literature here \cite{buerger2017multi}. However, we still want to investigate how the utilization of real microphones for identifying the \RIRs at the control points affects the reproduction performance. In addition to the inherent noise of microphones, \RIR measurements typically also suffer from positioning errors and mismatches between the characteristics of the individual microphones. According to the work by Cox et al. \cite{cox1986practical} and Teutsch \cite{teutsch2007modal}, these imperfections have a similar effect as spatially white noise. To assess the impact on the performance of \PM, \JPVM, and \JPVMmod, we therefore add spatially uncorrelated white Gaussian noise sequences to the clean free-field \RIRs corresponding to $\matgreensfunction$ in \cref{eq:pressure_and_velocity_minimization_problem}, where different \SNRs are considered. The free-field \RIRs are of length 128 samples and the \SNRs are determined based on the energies of the clean \RIRs and the noise sequences. The broadband reproduction error (MSE) in the bright zone and the broadband level difference $\Delta L$ obtained from averaging \cref{eq:MSE} and \cref{eq:level_difference}, respectively, are listed in \cref{tab:MSE_sensor_noise_combined} for the three approaches. The table contains results for \SNR values of $\unit[10]{dB}$, $\unit[20]{dB}$, $\unit[30]{dB}$, and $\unit[60]{dB}$, where only frequencies above $\unit[100]{Hz}$ are considered for evaluation. As expected, decreasing the \SNR results in an increase of the \MSE values and a reduction of the level difference. In case of \PM, the \MSE in the bright zone degrades by less than $\unit[0.5]{dB}$ when the \SNR is reduced from $\unit[60]{dB}$ to $\unit[10]{dB}$. Due to the exploitation of the pressure differences, both \JPVM and \JPVMmod are more susceptible to sensor noise, where the \MSE in the bright zone increases by more than $\unit[1]{dB}$ for both approaches. Nevertheless, the absolute \MSE values are always lowest when utilizing \JPVMmod. The impact of sensor noise on the level difference $\Delta L$ is most pronounced in case of \JPVMmod, where a reduction of the \SNR from $\unit[60]{dB}$ to $\unit[10]{dB}$ results in a degradation by more than $\unit[2.5]{dB}$. A slightly lower reduction of $\Delta L$ can be observed for \JPVM, whereas the decrease for \PM is below $\unit[1]{dB}$. However, \JPVMmod achieves the highest absolute level differences throughout the entire \SNR range. These results indicate that sensor noise should not be a severe problem in practice. It is worth noting that, albeit the average gain in level difference provided by \JPVMmod relative to \PM is only about $\unit[2]{dB}$ at most, the curves in \cref{fig:MSE_inside_and_contour,fig:level_differences} indicate that \JPVMmod introduces much less coloration, which is an important aspect when it comes the perceived reproduction quality.

\setlength\tabcolsep{-1pt}

\newcolumntype{C}[1]{>{\centering\let\newline\\\arraybackslash\hspace{0pt}}p{#1}}
\newlength{\mycolumnwidth}
\setlength{\mycolumnwidth}{1.4cm}
\newlength{\mygapwidth}
\setlength{\mygapwidth}{20pt}
\colorlet{tableheadcolor}{gray!25}
\renewcommand{\arraystretch}{1.2}
\begin{table}
	\centering
	\caption{Impact of sensor noise on the reproduction accuracy in the bright zone (MSE) and the level difference $\Delta L$ between the bright zone and the dark zone when using PM, JPVM, and \JPVMmod.}
	\vspace{10pt}
	\begin{tabular}{C{2\mycolumnwidth}C{\mycolumnwidth}C{\mycolumnwidth}C{\mygapwidth}C{\mycolumnwidth}C{\mycolumnwidth}C{\mygapwidth}C{\mycolumnwidth}C{\mycolumnwidth}C{\mygapwidth}}
		\rowcolor{black} & & & & & & & & & \\[-16pt]
		\rowcolor{tableheadcolor}
		& \multicolumn{2}{c}{\bfseries PM} 	& & \multicolumn{2}{c}{\bfseries JPVM} 	& & \multicolumn{2}{c}{\bfseries\JPVMmod} & \\
		\cellcolor{tableheadcolor} & \cellcolor{black} & \cellcolor{black} & \cellcolor{tableheadcolor} & \cellcolor{black} & \cellcolor{black} & \cellcolor{tableheadcolor} & \cellcolor{black}& \cellcolor{black} & \cellcolor{tableheadcolor}	\\[-16pt]
		\rowcolor{tableheadcolor} 
		\multirow{-2}{2cm}{\centering\bfseries SNR in dB}	& $\Delta L$ & MSE 		& & $\Delta L$ & MSE 		& & $\Delta L$ & MSE & \\ %\hline
		\rowcolor{black} & & & & & & & & & \\[-16pt] \rowcolor{white} 
		10 & 12.3 & -34.3 && 10.9 & -33.7 && 12.7 & -34.6 & \\ \hline
		20 & 13.0 & -34.7 && 12.7 & -34.9 && 14.8 & -35.8 & \\ \hline
		30 & 13.1 & -34.7 && 12.9 & -35.2 && 15.2 & -36.1 & \\ \hline
		60 & 13.1 & -34.8 && 13.0 & -35.2 && 15.3 & -36.2 & \\ \hline
	\end{tabular}
	\label{tab:MSE_sensor_noise_combined}
\end{table}

%\acresetall
% =========================================================================
\section{Conclusion}
\label{sec:conclusion}
% =========================================================================

In this paper, a recently proposed method for multizone sound reproduction, referred to as \JPVM, is analyzed and further developed to yield the improved \JPVMmod method. The core idea of \JPVM is to independently control the sound field in different zones by jointly optimizing the sound pressure and particle velocity vector on the surrounding contours. The analysis of \JPVM presented in this work utilizes spherical harmonics to describe the sound field and provides fundamental insights into its behavior. First of all, it illustrates that the optimization of the tangential component of the particle velocity vector is very prone to errors, and it may even degrade the reproduction performance. Therefore, \JPVMmod, an improved version of \JPVM is proposed which only considers the radial component of the particle velocity vector. As a result, the computational complexity is reduced, while increasing the reproduction performance, as is verified by simulation results. The simulations also reveal that the error of the sound pressure on the contour is not a reliable measure for inferring the error in the interior. Finally, it is shown that sensor noise does not have a significant impact on the reproduction performance of \JPVMmod despite relying on pressure differences.

%% before appendix (optional) and bibliography:
% =========================================================================
\begin{acknowledgments}
	The authors would like to express their sincere gratitude to thank Gary Elko, Thushara Abhayapala, Rudolf Rabenstein, Filippo Fazi, and Buye Xu for their helpful suggestions and fruitful discussions.
\end{acknowledgments}
% =========================================================================

\acresetall
% =========================================================================
\section*{Appendix: Relation between White Noise Gain and Loudspeaker Weight Energy}
% =========================================================================
In this section, we want to relate the \LWE \cite{betlehem2012sound} to the \WNG \cite{cox1986practical}, which is a well-known and commonly used robustness measure in beamforming. The \WNG is defined as the \SNR obtained for a particular `target direction' in the presence of spatially white transducer noise of unit variance. In the context of multizone sound rendering, not only the direction, but also the distance is crucial, such that the target direction becomes a `target position'. In fact, it is only meaningful to compute the \WNG for target positions $\xxbright$ in the bright zone $\brightzone$, as the desired signal in the dark zone $\darkzone$ is typically zero which, in the ideal case, would imply an \SNR of $\unit[-\infty]{dB}$.

To establish a relation between the \WNG and the \LWE, let us consider the optimization problem for pressure matching \cref{eq:pressure_matching_aim} with an additional constraint on the LWE, 
\begin{equation}
\min\limits_{\mat{w(\omega)}}  \norm{\matgreensfunction(\omega) \mat{w(\omega)} - \vectransferfunctiondes(\omega)}^2 \quad \text{s.~t.}\quad \norm{\mat{w}(\omega)}^2 \leq \alpha,
\label{eq:constraint_optimization_problem_appendix}
\end{equation}
where $\alpha$ is the upper bound for the \LWE. The target sound field in the bright zone is chosen as
\begin{equation}
\Gdes\left(\xxbright,\omega\right) = \frac{1}{4 \pi \norm{\yy - \xxbright}} \e{-\jj k \norm{\yy - \xxbright}},
\label{eq:target_sound_field_appendix}
\end{equation}
which corresponds to the sound pressure that would be obtained with a single loudspeaker located at position $\yy$. 
For simplicity, we assume in the following that the bright zone is located at the center of a circular loudspeaker array of radius $r_0$. Furthermore, it is assumed that $\norm{\yy}=r_0$ and that the bright zone is sufficiently small relative to the array radius $r_0$, \ie, the distance-dependent attenuation of the spherical waves emitted by the loudspeakers is approximately identical for all $\xxbright$. Given that the optimization problem in \cref{eq:constraint_optimization_problem_appendix} is solved reasonably well, the overall transfer function of the reproduction system \wrt the sound pressure at the target positions in the bright zone may be approximated as
\begin{equation}
\vecgreensfunction\tran\big(\xxbright,\omega\big) \vecrenderingfilt(\omega) \approx \frac{1}{4\pi r_0} \e{-\jj k \norm{\yy - \xxbright}}. 
\label{eq:approx_target_transfer_function_bright_zone}
\end{equation}
Note that, due to the assumption that the zone size is small relative to $r_0$, the level within the bright zone is approximately constant.

To specify the \WNG, we also need to determine the noise signal $N\big(\xxbright,\omega\big)$ at the target position $\xxbright$ as resulting from spatially white transducer noise. Denoting the spatially white noise signal originating from the $l$-th loudspeaker as $N_l(\omega)$, $l=1,\ldots,\NL$, the resulting noise signal $N\big(\xxbright,\omega\big)$ is given by (\cf \cref{eq:reproduced_pressure})
\begin{equation}
\begin{split}
N \big(\xxbright,\omega \big) 	&= \sum\limits_{l=1}^{\NL} \greensfunction\big(\xxbright|\xxL_l,\omega\big) W_l(\omega) N_l(\omega) \\
&= \mat{w}\tran(\omega) \mat{n}\big(\xxbright,\omega\big),
\label{eq:spatially_white_noise_at_target_point}
\end{split}
\end{equation}
where $\mat{n}\big(\xxbright,\omega\big) = \big[\greensfunction\big(\xxbright|\xxL_1,\omega\big) N_1(\omega),\ldots, \greensfunction\big(\xxbright|\xxL_\NL,\omega\big) N_\NL(\omega)\big]\tran$ is a vector capturing the contributions of each loudspeaker to the noise field at $\xxbright$.

According to its definition, the \WNG can be expressed as
\begin{equation}
\begin{split}
\WNG\big(\xxbright,\omega\big) 	&= \frac{ \abs{\vecgreensfunction\tran \big(\xxbright,\omega\big) \mat{w(\omega)} }^2 }{ \expval{N \big(\xxbright,\omega\big) 
		N^*\big(\xxbright,\omega\big)} } \\
&\approx \frac{\frac{1}{(4\pi r_0)^2}}{ \mat{w}\tran(\omega) \underbrace{\expval{ \mat{n}\big(\xxbright,\omega\big) \mat{n}\herm\big(\xxbright,\omega\big)}}_{\mat{R}(\xxbright,\omega)} \mat{w}^*(\omega) } ,
\label{eq:WNG_definition}
\end{split}
\end{equation}
where $\expval{\cdot}$ is the expectation operator and the superscript $(\cdot)^*$ denotes complex conjugation. As the noise signals $N_l$ are spatially white, the correlation matrix $\mat{R}$ is diagonal and contains the squared magnitudes of the individual Green's functions. Due to the assumption that these magnitudes are approximately equal for all target positions, the correlation matrix can be approximated as 
\begin{equation}
\mat{R}(\xxbright,\omega) \approx \frac{1}{(4 \pi r_0)^2} \mat{I}_\NL,
\label{eq:correlation_matrix}
\end{equation}
with $\mat{I}_\NL$ being an identity matrix of dimensions $\NL \times \NL$. 
Inserting \cref{eq:correlation_matrix} into \cref{eq:WNG_definition} finally yields 
\begin{equation}
\WNG\big(\xxbright,\omega\big) \approx \frac{\frac{1}{(4\pi r_0)^2}}{ \frac{1}{(4\pi r_0)^2} \mat{w}\tran(\omega) \mat{w}^*(\omega) } = \frac{1}{\norm{\mat{w}(\omega)}^2}.
\label{eq:relation_WNG_LWE}
\end{equation}
That is, the \WNG can be approximated by the inverse of the \LWE. Note that \cref{eq:relation_WNG_LWE} is only valid if the target sound field for the bright zone is specified such that the distance-dependent attenuation resulting from the Green's functions is considered. If the target sound field in \cref{eq:target_sound_field_appendix} was defined with a magnitude of one, the \WNG in \cref{eq:relation_WNG_LWE} would be increased by a factor of $(4\pi r_0)^2$. It shall finally be noted that, even though the distances between all loudspeakers and target positions are not identical in practice, \cref{eq:relation_WNG_LWE} can still be used to obtain an approximate relation between the \WNG and the \LWE.

% ================================================================================================
\bibliography{refs}
\bibliographystyle{ieeetr} %jasa

% ================================================================================================

\end{document}